\documentclass[twocolumn]{aastex63}
\usepackage{natbib}
\usepackage{color}
\usepackage{mathtools}
\usepackage{hyperref} 

\def    \apjl  		{\rm {ApJL}}

\def    \apjl  		{\rm {ApJL}}

\def	\cm		{\,{\rm {cm}}}
\def	\K		{\,{\rm K}}
\def	\g		{\,{\rm {g}}}
\def	\mum	{\,{\mu \rm{m}}}

\def \bea {\begin{eqnarray}}
\def \ena {\end{eqnarray}}

\def	\B	{{\rm B}}

\def	\C	{{\rm C}}

\def	\cm	{\,{\rm cm}}
\def	\d	{{\rm d}}

\def	\D	{{\rm D}}

\def	\erg	{\,{\rm erg}}

\def	\g	{\,{\rm g}}
\def	\gas	{\,{\rm gas}}
\def	\G	{{\rm G}}

\def	\H	{{\rm H}}

\def	\N	{{\rm N}}

\def	\s	{\,{\rm s}}

\def	\V	{{\rm V}}

\def	\rad	{{\rm rad}}

\def	\H	{{\rm H}}
\def	\C	{{\rm C}}
\def	\O	{{\rm O}}
\def	\OH	{{\rm OH}}
\def	\CH	{{\rm CH}}
\def	\CO	{{\rm CO}}
\def	\HCO	{{\rm HCO}}
\def	\CHO	{{\rm CHO}}

\def    \gas     	{{\rm gas}}

\begin{document}
\shorttitle{Ro-thermal hopping and segregation}
\shortauthors{Thiem Hoang}
\title{Chemistry on rotating grain surfaces: ro-thermal hopping and segregation of molecules in ice mantles}
\author{Thiem Hoang}
\affil{Korea Astronomy and Space Science Institute, Daejeon 34055, Republic of Korea}
\affil{University of Science and Technology, Korea, (UST), 217 Gajeong-ro Yuseong-gu, Daejeon 34113, Republic of Korea}

\begin{abstract}
Grain surfaces play a central role in the formation and desorption of molecules in space. To form molecules on a grain surface, adsorbed species trapped in binding sites must be mobile and migrate to adjacent sites. Thermal hopping is a popular mechanism for the migration of adsorbed species when the grain surface is warmed up by stellar radiation. However, previous studies disregarded the fact that grains can be spun-up to suprathermal rotation by radiative torques (RATs) during grain heating process. To achieve an accurate model of surface astrochemistry, in this paper, we study the effect of grain suprathermal rotation by RATs on thermal hopping of adsorbed species on icy grain mantles. We find that centrifugal force due to grain suprathermal rotation can increase the mobility of radicals on/in the ice mantle compared to the prediction by thermal hopping, and we term this mechanism ro-thermal hopping. The rate of ro-thermal hopping depends both on the local radiation energy density (i.e., grain temperature) and gas density, whereas thermal hopping only depends on grain temperature. We calculate the decrease in grain temperature required by ro-thermal hopping to produce the same hopping rate as thermal hopping and find that it increases with increasing the diffusion energy and decreasing the gas density. We finally study the effect of grain suprathermal rotation on the segregation of ice mixtures and find that ro-thermal segregation of CO$_2$ from H$_2$O-CO$_2$ ices can occur at much lower temperatures than thermal segregation reported by experiments. Our results indicate that grain suprathermal rotation can enhance mobility, formation, desorption, and segregation of molecules in icy grain mantles.

\end{abstract}

\section{Introduction}
Water is essential for life, and complex organic molecules (COMs, having more than six atoms) are considered the building block of life. To date, more than 200 different molecules, including water and COMs, were detected in the gas phase of outer space, including the interstellar medium (ISM), star-forming regions, and protoplanetary disks. The remaining question is where and how COMs form and return into the gas (see e.g., \citealt{Caselli:2012fq}, \citealt{2013RvMP...85.1021T}; \citealt{2017SSRv..212....1C}; \citealt{2018IAUS..332....3V} for reviews). 

Grain surfaces and ice mantles are believed to play a crucial role in the formation of molecules, including H$_{2}$, H$_{2}$O, and COMs (see e.g., \citealt{Herbst:2009go}). Grain surface chemistry in general involves four main physical processes: (1) accretion of gas atoms/molecules to the grain surface, (2) mobility of adsorbed species on or in the ice mantle, (3) probability to form molecules upon collisions, and (4) desorption of newly formed molecules from the grain surface (see \citealt{2018IAUS..332....3V}). 

The rate of surface chemical reactions depends on the mobility of adsorbed species on the grain surface (\citealt{1972ApJ...174..321W}; \citealt{1992ApJS...82..167H}). Because species physically adsorbed to the grain surface are trapped in a potential minimum, they cannot freely migrate over the grain surface. Thus, to enable chemical reaction, adsorbed species must receive some kinetic energy to overcome the diffusion potential barrier ($E_{\rm diff}$) such that they can migrate from one binding site to adjacent ones. Thermal hopping due to heating of ice mantles by stellar radiation is a popular mechanism to increase the mobility of adsorbed simple species (hereafter primary radicals) and trigger surface chemical reactions (e.g., \citealt{1992ApJS...82..167H}). Primary radicals have low $E_{\rm diff}$ such that the first stage of chemical reaction on the grain surface is expected to occur during the warming up phase by star formation when the grain temperature is increased from $10\K$ to $T_{d}>20\K$, forming secondary radicals (see \citealt{Herbst:2009go} and \cite{2018IAUS..332....3V} for recent reviews). Subsequently, when the ice mantle is heated to higher temperatures, secondary radicals can react with primary radicals to form more complex molecules, including COMs. For low temperatures of $T_{d}<15\K$, the diffusion due to the tunneling quantum effect is more important than thermal hopping (\citealt{1972ApJ...174..321W}). 

Star-forming regions and photodissociation regions (PDRs) are also known to contain rich chemistry (see e.g. \citealt{1997ARA&A..35..179H}). Most of COMs are observed toward hot cores/corinos around young stars. The popular scenario to form COMs in hot cores/corinos involves three phases, cold, warm, and hot phases \citep{Herbst:2009go}. During the initial cold phase, COMs may first be formed in the cold molecular core (\citealt{2016ApJ...830L...6J}) but are frozen in the icy grain mantle during the cloud collapse process (zeroth-generation species). During the {\it warm phase} upon star formation, the ice mantle is warmed up from $\sim 10$ K to $\sim 100$ K by protostellar radiation, which increases the mobility of simple molecules frozen in the ice mantle and finally form COMs (first-generation species; see e.g., \citealt{2008ApJ...682..283G}). During the {\it hot phase} where icy grain mantles are heated to $T_{d}\sim 100 - 300$ K, thermal sublimation of ice mantles (\citealt{1987ApJ...315..621B}; \citealt{1988MNRAS.231..409B}; \citealt{Bisschop:2007cu}) can release radicals (CH$_3$OH, NH$_3$), which trigger gas-phase chemistry at high temperatures and form in-situ COMs (second-generation species; see \citealt{1992ApJ...399L..71C}). Therefore, the warm phase plays a key role for formation of COMs. 

Previous studies on mobility, formation, and desorption of molecules assumed that grain surfaces are at rest, which is contrary to the fact that grains are rapidly rotating due to collisions with gas atoms and interstellar photons (\citealt{1998ApJ...508..157D}; \citealt{Hoang:2010jy}). In particular, interstellar dust grains are known to be rotating suprathermally (i.e., grain angular velocities well above their thermal values, \citealt{1979ApJ...231..404P}). Suprathermal rotation is indeed required for efficient alignment of grains with magnetic fields (\citealt{2016ApJ...831..159H}) in order to explain starlight polarization and far-IR/submm polarized dust emission (see \citealt{Andersson:2015bq} and \citealt{LAH15} for reviews). Modern understanding of dust astrophysics reveals that dust grains of irregular shapes can be spun-up to suprathermal rotation (with angular velocities larger than thermal velocity) by radiative torques induced by the illumination of an anisotropic radiation field (\citealt{1996ApJ...470..551D}; \citealt{2007MNRAS.378..910L}; \citealt{Hoang:2008gb}; \citealt{2009ApJ...695.1457H}; \citealt{Herranen:2019kj}). Subject to intense radiation of protostars, dust grains are simultaneously heated to high temperatures and spun-up to extremely fast rotation. Although the grain heating effect is well-known to be important for mobility and desorption of molecules on the grain surface (see e.g., \citealt{1972ApJ...174..321W}; \citealt{1982A&A...114..245T}), the effect of grain suprathermal rotation is neglected. Moreover, grains can also rotate suprathermally as a result of mechanical torques induced by an anisotropic gas flow (\citealt{2007ApJ...669L..77L}; \citealt{2018ApJ...852..129H}). 

\cite{Hoang:2019td} first studied the effect of suprathermal rotation on the desorption of icy grain mantles and introduced a mechanism called rotational desorption. The proposed mechanism can disrupt the ice mantle into tiny fragments, subsequently molecules rapidly evaporate from the tiny ice fragments. Later, \cite{Hoang:2019wra} considered the effect of suprathermal rotation on thermal sublimation of molecules from intact ice mantles. The authors find that centrifugal potential energy can lower the potential barrier for desorption defined by binding energy $E_{b}$ such that molecules can desorb from the ice mantle at temperatures much below their sublimation threshold.

Since adsorbed species are trapped in a potential well with an energy barrier of $E_{\rm diff}\ll E_{b}$, the centrifugal potential energy due to rotation can reduce the diffusion barrier and assist thermal diffusion/hopping of species over the grain surface. As a result, the mobility of species is enhanced and the reaction rate is increased accordingly. Within our effort to investigate the effect of grain rotation on surface chemistry, in this paper, we study the effect of suprathermal rotation on thermal hopping of adsorbed species on/in the ice mantle.


The structure of the paper is as follows. In Section \ref{sec:theory} we consider the effect of grain rotation on thermal hopping of adsorbed atoms/molecules and introduce rotational-thermal hopping on the rotating grain surface. In Section \ref{sec:ro-RAT} we apply our ro-thermal hopping mechanism for the case of grain suprathermal rotation spun-up by radiative torques. In Section \ref{sec:discuss} we discuss the effect of grain suprathermal rotation on the segregation of ice mixtures and environments where ro-thermal hopping and segregation are important. The main findings are summarized in Section \ref{sec:sum}.

\section{Ro-thermal hopping on rotating grain surfaces}\label{sec:theory}
We first introduce the physical model of ro-thermal hopping of adsorbed molecules on rotating grain surfaces.
\subsection{Icy grain model}
\begin{figure}
\includegraphics[scale=0.45]{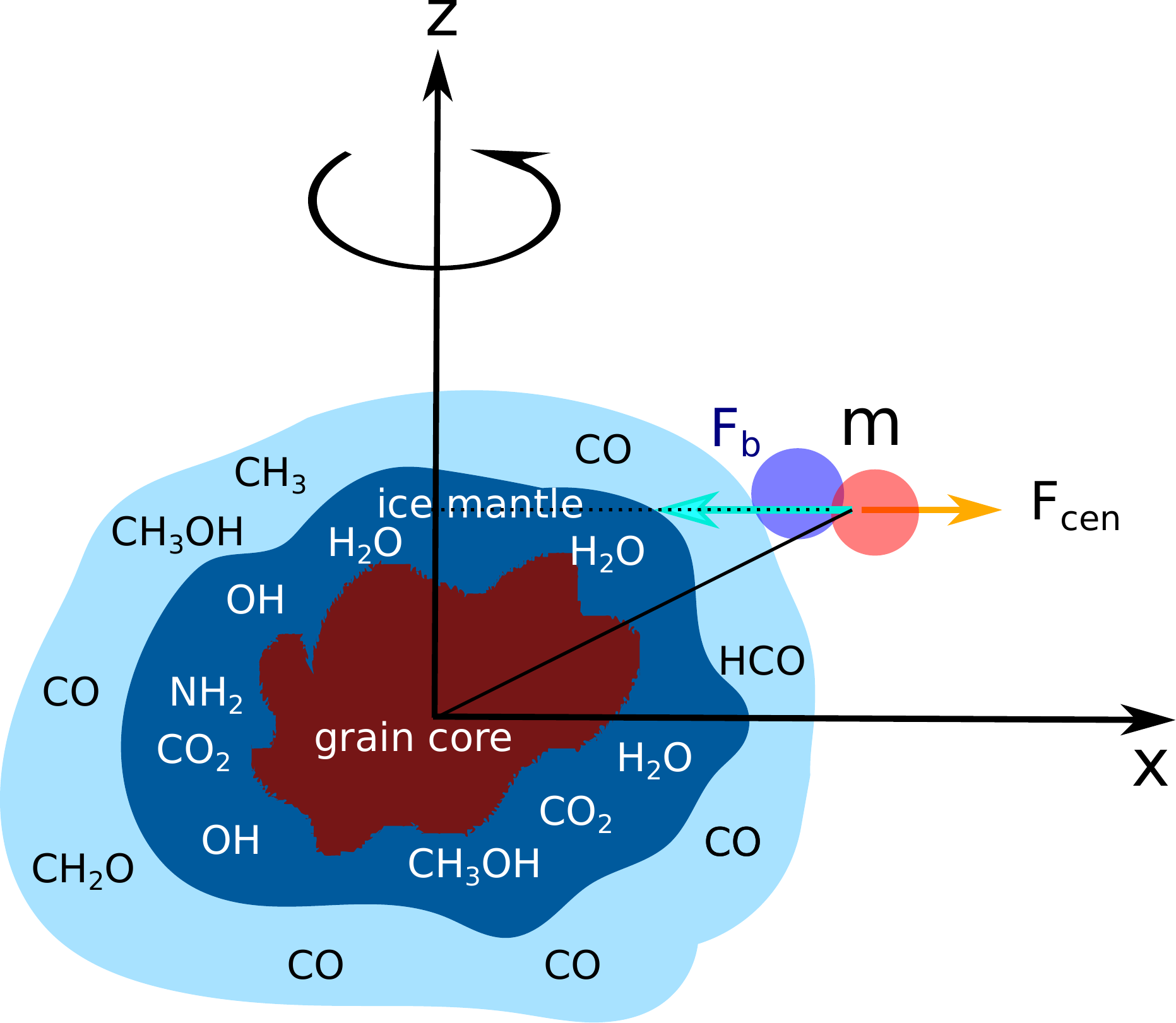}
\caption{A schematic illustration of a core-mantle grain of irregular shape spinning around the $z$-axis. The silicate core is assumed to be compact, which is covered with a H$_2$O-dominated and CO-dominated ice mantles. A molecule of mass $m$ on the ice surface experiences the binding force $F_{b}$ and centrifugal force $F_{\rm cen}$ which are in opposite directions.}
\label{fig:grainmodel}
\end{figure}

We consider a grain model consisting of a silicate core covered with an ice mantle. This grain model is expected for grains in star-forming regions because the ice mantle forms on the grain surface during the cold phase where the gas temperature is as low as $10\K$ (\citealt{1983Natur.303..218W}). Spectral absorption features of H$_{2}$O and CO ice are highly polarized (\citealt{1996ApJ...465L..61C}; \citealt{2008ApJ...674..304W}), which requires the icy grain mantles to have non-spherical shape and be aligned with magnetic fields (see \citealt{LAH15} for a review). Nevertheless, we can assume that the grain shape can be described by an equivalent sphere of the same volume with effective radius $a$. Figure \ref{fig:grainmodel} shows the illustration of the icy grain mantle surrounding a grain core. 

\subsection{Accretion of gas species to grain surfaces}
The accretion rate of gas species onto a grain surface is given by
\bea
k_{\rm acc}=s\langle v\rangle 4\pi a^{2},\label{eq:kacc}
\ena
where $a$ is the grain radius, $s$ is the sticking coefficient, and $\langle v\rangle$ is the mean thermal velocity of the species. 

The time interval between two successive landings of a molecule on the grain surface is $t_{\rm acc}^{-1}=k_{\rm acc}n(A)$ where $n(A)$ is the gas density of species A (see, e.g., \citealt{1998ApJ...495..309C}).


When a gas species arrives on the grain surface, it can be weakly attached to the surface due to van der Waals force. This is called physiochemical sites, and the binding energy is low about $0.01$ eV. When the distance to the surface is reduced, the chemical bonding starts to act and the binding energy is larger up to 1 eV. For grains in star-forming regions, only physiochemical sites exist on the icy grain mantle.

Let $N_{s}=4\pi a^{2}\sigma_{s}$ be the total number of catalytic sites on the grain surface of radius $a$, where $\sigma_{s}$ is the surface density of active sites. Experimental value gives $\sigma_{s}\sim 2\times 10^{14}\cm^{-2}$, which yields $N_{s}=10^{5}$ for $a=0.1\mum$.

\subsection{Ro-thermal hopping}
\begin{figure}
\includegraphics[scale=0.4]{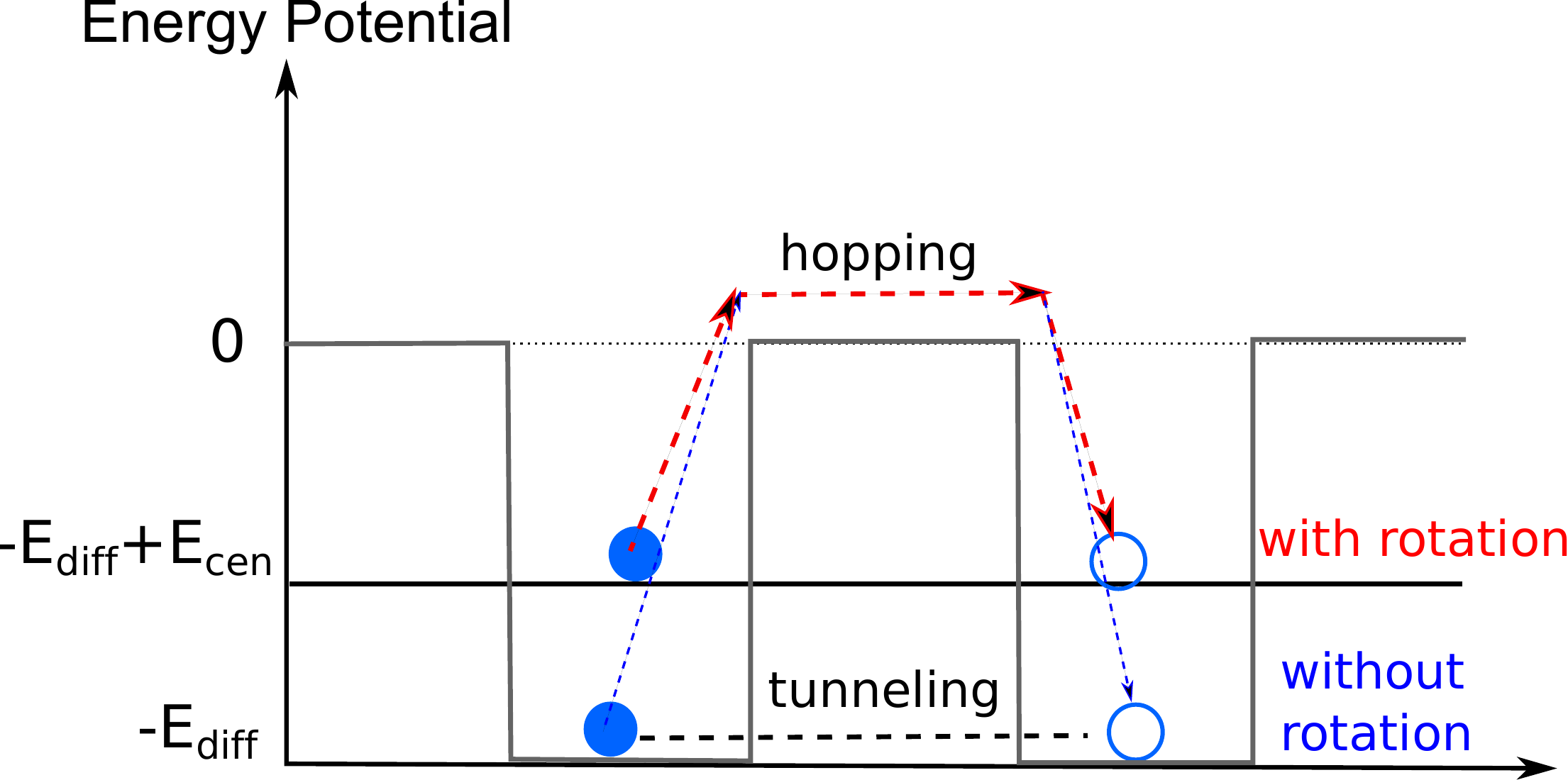}
\caption{Schematic illustration for the diffusion of adsorbed species from one catalytic site to adjacent site on the grain surface by thermal hopping and tunneling effect with and without grain rotation. Centrifugal force due to grain rotation reduces the diffusion barrier and enhances the diffusion of adsorbed species.}
\label{fig:diffusion}
\end{figure}

Adsorbed species on the grain surface can migrate on the grain surface through hopping and tunneling (see e.g., \citealt{1982A&A...114..245T}). Figure \ref{fig:diffusion} describes the diffusion of adsorbed species from a potential well (surface site) to an adjacent one due to thermal hoping and tunneling. The tunneling effect is important when the surface temperature is low (i.e.,$T_{d} <10$ K), and thermal hopping dominates when the grain is heated to higher temperatures of $T_{d}>10$ K (\citealt{Tielens:2007wo}).

Previous studies disregarded the effect of grain rotation on thermal hopping of adsorbed species on the surface. Thereby, the timescale of thermal hopping is described by (\citealt{1972ApJ...174..321W}):
\bea
t_{hop,0}=\nu_{0}^{-1}\exp\left(\frac{E_{\rm diff}}{kT_{d}} \right),\label{eq:khop_0}
\ena
where $E_{\rm diff}$ is the potential energy barrier between adjacent surface potential energy wells of accreted species. Typically, the potential energy barrier $E_{\rm diff}\approx 0.05-0.5E_{b}$ where $E_{b}$ is the binding energy of the molecule to the grain surface (\citealt{1992ApJS...82..167H}). Here, the characteristic frequency is given by
\bea
\nu_{0}=\left(\frac{2\sigma_{s}E_{b}}{\pi^{2}m}\right)^{2},
\ena
where $m$ is the mass of adsorbed species (\citealt{1982A&A...114..245T}). Thus, one has $\nu_{0}\sim 10^{12}-10^{13}\s^{-1}$.

Taking into account the effect of grain suprathermal rotation, the diffusion barrier potential is reduced. Thus, the timescale of {\it ro-thermal hopping} is reduced to
\bea
t_{hop,rot}&=&\nu_{0}^{-1}\exp\left(\frac{E_{\rm diff}-m\langle \phi_{\rm cen}\rangle}{kT_{d}} \right)\nonumber\\
&=& t_{hop,0}\exp\left(-\frac{m\langle \phi_{\rm cen}\rangle}{kT_{d}} \right),
\label{eq:thop}
\ena
where $\langle \phi_{\rm cen}\rangle$ is the centrifugal potential as averaged over the grain surface (\citealt{Hoang:2019wra}):   
\bea
\langle \phi_{\rm cen}\rangle=\frac{\omega^{2}a^{2}}{3}.\label{eq:phi_cen}
\ena

The rate of thermal hopping and ro-thermal hopping is given by $k_{hop,i}=1/t_{hop,i}$ for $i=0$ and $rot$ for non-rotating and rotating grains. Therefore, Equation (\ref{eq:thop}) can be rewritten as
\bea
k_{hop,rot}=k_{hop,0}\exp\left(\frac{m\langle \phi_{\rm cen}\rangle}{kT_{d}} \right).\label{eq:khop}
\ena

Equation (\ref{eq:khop}) reveals that the hopping rate increases with increasing the mass of the adsorbed species. Therefore, heavier species are more mobile than light species if the centrifugal potential $\phi_{\rm cen}$ is comparable to the diffusion barrier potential. This is contrary to the case of non-rotating surfaces where light species are more mobile than heavy ones due to their lower $E_{\rm diff}$.

\begin{table}
\begin{center}
\caption{Binding energies of selected radicals on an ice surface.}\label{tab:Ebind}
\begin{tabular}{l l l l} \hline\hline
{Molecules} & {$E_{b}/k$ (K)$^a$ } & {Molecules} & {$E_{b}$/k (K)$^a$ }\cr
\hline\\

$\rm CO$ & 1150 & $\rm CH_3CO$ & 2325 \cr
$\rm CH_{3}$ & 1175 & $\rm COCHO $& 2750 \cr
$\rm CH_{4}$ & 1300 & $\rm HNCO $& 2850\cr
$\rm NH$ & 2380  & $\rm CH_3NH$ & 3554\cr
$\rm CH_{3}O$ & 2500 &  $\rm CH_3OCO$ & 3600\cr
$\rm OH$ & 2850 &  $\rm HNCHO$ & 3978\cr
$\rm NH_{2}$ & 3960 & $\rm N_2H_2 $ & 4756\cr
$\rm CH_{2}OH$ & 5080 & $\rm CH_3ONH $& 4828  \cr
$\rm H_{2}O$ & 5700 &  $\rm NH_2CO$ & 4506 \cr
$\rm CO_{2}$ & 2575 & $\rm COOH $ & 5120 \cr
 
\cr
\hline
\multicolumn{3}{l}{$^a$~See Tables 1 and 2 in \cite{2008ApJ...682..283G}}\cr
\cr
\end{tabular}
\end{center}
\end{table}

Table \ref{tab:Ebind} shows the binding energy of primary and secondary radicals where $E_{b}=2E_{\rm diff}$ using the data of $E_{\rm diff}$ from Table 1 and 2 from \cite{2008ApJ...682..283G}.

\subsection{Temperatures of ro-thermal hopping}
Let $T_{\rm hop,0}$ be the hopping temperature of adsorbed species on the grain at rest, i.e., $\omega=0$. The hopping temperature on a rotating grain is denoted by $T_{\rm hop,rot}$.

To quantify the effect of grain rotation on thermal hopping, we compare the grain temperature that is required to produce the same hopping rate from a non-rotating grain which corresponds to $t_{\rm hop,0}(T_{\rm hop,0})=t_{\rm hop,rot}(T_{\rm hop,rot})$. Thus, one obtains
\bea
T_{\rm hop,rot}=\left(1-\frac{m\langle \phi_{\rm cen}\rangle}{E_{\rm diff}} \right)T_{\rm hop,0}.\label{eq:Thop_rot}
\ena

The effect of grain rotation reduces the sublimation temperature as given by
\bea
\frac{T_{\rm hop,0}-T_{\rm hop,rot}}{T_{\rm hop,0}}&=&\left(\frac{m\langle \phi_{\rm cen}\rangle}{E_{\rm diff}}\right)=\left(\frac{ma^{2}\omega^{2}}{3E_{\rm diff}}\right)\label{eq:dThop}\\
&\simeq& 0.15a_{-5}^{2}\left(\frac{\omega}{10^{9}\s^{-1}}\right)^{2}\left(\frac{m}{m(\rm CO_{2})}\right)\left(\frac{1000\K}{(E_{\rm diff}/k)}\right),\nonumber
\ena
where $a_{-5}=a/(10^{-5}\cm)$.

\subsection{Surface chemical reaction}

Let $N(A)$ and $N(B)$ be the number of $A$ and $B$ species on the grain surface. Both species move on the grain surface and react upon encounters, which is known as the Langmuir-Hinshelwood mechanism. The reaction rate for species A is given by (see e.g., \citealt{Herbst:2005bd}; \citealt{2017SSRv..212....1C})
\bea
\frac{dN(A)}{dt}=- \kappa k_{AB}N(A)N(B),\label{eq:rate}
\ena
where $k_{AB}$ is the reaction rate coefficient between $A$ and $B$ species as given by
\bea
k_{AB}=\frac{k_{hop}^{A}+k_{hop}^{B}}{N_{s}},
\ena
and $\kappa$ is the probability of reactions upon encounters. Here the rate at which species scan the entire surface is $k_{\rm scan}=k_{\rm hop}/N_{s}$, i.e., the timescale to scan the entire surface of $N_{s}$ sites is equal to $t_{\rm scan}=N_{s}t_{\rm hop}$ where $t_{\rm hop}$ is the hopping time from one site to an adjacent site.


Including both the accretion and desorption of species into Equation (\ref{eq:rate}), one obtains the net reaction rate:
\bea
\frac{dN(A)}{dt}=k_{\rm acc}^{A}n(A)- \kappa k_{AB}N(A)N(B)-k_{\rm des}^{A}N(A),\label{eq:netrate}
\ena
where $n(A)$ is the number density of species A in the gas phase, $k_{\rm des}$ is the desorption rate of species A. 

Following \cite{Hoang:2019wra}, the rate of ro-thermal desorption in the presence of grain rotation is given by
\bea
k_{\rm des,rot}=k_{\rm des,0}\exp\left(\frac{m\langle \phi_{\rm cen}\rangle}{kT_{d}} \right),\label{eq:kdes_rot}
\ena
where $k_{\rm des,0}$ is the thermal sublimation rate without grain rotation as given by
\bea
k_{\rm des,0}=\nu_{0}\exp\left(-\frac{E_{b}}{kT_{d}} \right).
\ena

\section{Ro-thermal hopping by radiative torques}\label{sec:ro-RAT}
Now we apply our theory in the previous section for the case in which grains are spun-up to suprathermal rotation by radiative torques.

\subsection{Centrifugal potential due to radiative torques}
Consider dust grains of irregular shape illuminated by a radiation field of anisotropy degree $\gamma$, mean wavelength $\bar{\lambda}$, and radiation energy density $u_{\rm rad}$. Such irregular grains experience radiative torques (RATs) due to differential scattering and absorption by anisotropic radiation, which can spin-up the grains to suprathermal rotation (\citealt{1996ApJ...470..551D}; \citealt{Hoang:2008gb}; \citealt{2016ApJ...831..159H}). The grain rotation is damped due to collisions with gas species and emission of infrared photons (see e.g., \citealt{Hoang:2019da}). As a result, the maximum rotation rate spun-up by RATs for an irregular grain of effective size $a$ is given by (\citealt{Hoang:2019bi}):
\bea
\omega_{\rm RAT}&\simeq &9.6\times 10^{9}\gamma a_{-5}^{0.7}\bar{\lambda}_{0.5}^{-1.7}\nonumber\\
&\times&\left(\frac{U}{n_{\H}T_{2}^{1/2}}\right)\left(\frac{1}{1+F_{\rm IR}}\right)\rad\s^{-1}~~~\label{eq:omega_RAT1}
\ena
for grains with $a\lesssim a_{\rm trans}=\bar{\lambda}/1.8$, and
\bea
\omega_{\rm RAT}&\simeq &1.8\times 10^{11}\gamma a_{-5}^{-2}\bar{\lambda}_{0.5}\nonumber\\
&&\times\left(\frac{U}{n_{\H}T_{2}^{1/2}}\right)\left(\frac{1}{1+F_{\rm IR}}\right)\rad\s^{-1}~~~\label{eq:omegaRAT2}
\ena
for grains with $a> a_{\rm trans}$ where $a_{\rm trans}$ denotes the grain size at which the RAT efficiency changes between the power law and flat stages (see \citealt{2007MNRAS.378..910L}; \citealt{Herranen:2019kj}). Here, $\bar{\lambda}_{0.5}=\bar{\lambda}/(0.5\mum)$, $n_{\H}$ is the hydrogen density, $T_{\gas}$ is the gas temperature with $T_{2}=T_{\gas}/100\K$, $F_{\rm IR}$ is the dimensionless parameter describing the grain rotational damping by infrared emission (\citealt{1998ApJ...508..157D}; \citealt{Hoang:2010jy}), and $U=u_{\rad}/u_{\rm ISRF}$ with $u_{\rm ISRF}=8.64\times 10^{-13}\erg\cm^{-3}$ being the radiation energy density of the standard interstellar radiation field (ISRF) in the solar neighborhood (\citealt{1983A&A...128..212M}; \citealt{Hoang:2019da}). 

Equations (\ref{eq:omega_RAT1}) and (\ref{eq:omegaRAT2}) reveal that the grain rotation rate depends on the physical term $U/n_{\H}T_{\gas}^{1/2}$ and $F_{\rm IR}$, which reflects the balance between spin-up by RATs and damping by gas collisions and IR emission.

Plugging $\omega_{\rm RAT}$ into Equation (\ref{eq:phi_cen}), one obtains the centrifugal potential due to grain rotation as follows:
\bea
m\langle \phi_{\rm cen}\rangle&\simeq&0.02\gamma^{2} a_{-5}^{3.4}\bar{\lambda}_{0.5}^{-3.4}\left(\frac{m}{m_{\rm CO}}\right)\nonumber\\
&\times&\left(\frac{U}{n_{\H}T_{2}^{1/2}}\right)^{2}\left(\frac{1}{1+F_{\rm IR}}\right)^{2}~\rm eV\label{eq:phi_small}
\ena
for $a\lesssim a_{\rm trans}$ and
\bea
m\langle \phi_{\rm cen}\rangle&\simeq&6.0\gamma^{2} a_{-5}^{-2}\bar{\lambda}_{0.5}^{2}\left(\frac{m}{m_{\rm CO}}\right)\nonumber\\
&\times&\left(\frac{U}{n_{\H}T_{2}^{1/2}}\right)^{2}\left(\frac{1}{1+F_{\rm IR}}\right)^{2} \rm eV\label{eq:phi_big}
\ena
for $a>a_{\rm trans}$.

The centrifugal potential increases rapidly with the grain size as $a^{3.4}$ until $a=a_{\rm trans}$ (Eq.\ref{eq:phi_small}) and with the radiation strength as $U^{2}$. Thus, the centrifugal potential is important for strong radiation fields. 

Using the centrifugal potentials (Eqs. \ref{eq:phi_small} and \ref{eq:phi_big}) one can calculate the rate of ro-thermal hopping (Eq. \ref{eq:khop}) and the temperature threshold for ro-thermal hopping (Eq. \ref{eq:Thop_rot}).

Note that individual molecules can be directly ejected by centrifugal forces when the rotational rate is sufficiently high. Following  \cite{Hoang:2019wra}, the critical radiation strength for the direct ejection is given by
\bea
U_{\rm ej}\simeq 0.8n_{\H}T_{2}^{1/2}(1+F_{\rm IR})\frac{\lambda_{0.5}^{1.7}}{\gamma a_{-5}^{1.7}}\left(\frac{(E_{b}/k)}{1300\K}\frac{m_{\rm CO}}{m}\right)^{1/2}~\label{eq:Uej}
\ena
for $a\lesssim a_{\rm trans}$, and 
\bea
U_{\rm ej}\simeq 0.04n_{\H}T_{2}^{1/2}(1+F_{\rm IR})\frac{\lambda_{0.5}^{1.7} a_{-5}}{\gamma}\left(\frac{(E_{b}/k)}{1300\K}\frac{m_{\rm CO}}{m}\right)^{1/2}
\ena
for $a> a_{\rm trans}$.

\subsection{Numerical Results}
\subsubsection{Ro-thermal hopping versus thermal hopping}
For numerical results, we will consider a wide range of parameter space with gas density $n_{\H}\sim 10^{2}-10^{6}\cm^{-3}$ and the radiation strength $U\sim 10^{2}-10^{6}$, which reflects various astrophysical environments from PDRs, to circumstellar disks, to hot cores/corinos of star-forming regions. The corresponding grain temperature is estimated by $T_{d}\simeq 16.4 a_{-5}^{-1/15}U^{1/6}$ for silicate grains (\citealt{2011piim.book.....D}), which is appropriate for silicate core-ice mantle grains considered in this paper. Assuming a thermal equilibrium between gas and dust, so that $T_{\rm gas}=T_{d}$.

We consider both simple molecules (primary radicals) frozen onto the ice mantle and products resulting from the reaction of primary radicals on the grain surface (hereafter secondary radicals, see Table \ref{tab:Ebind}). We assume $E_{\rm diff}=0.5E_{b}$ for our calculations. 

\begin{figure*}
\includegraphics[scale=0.5]{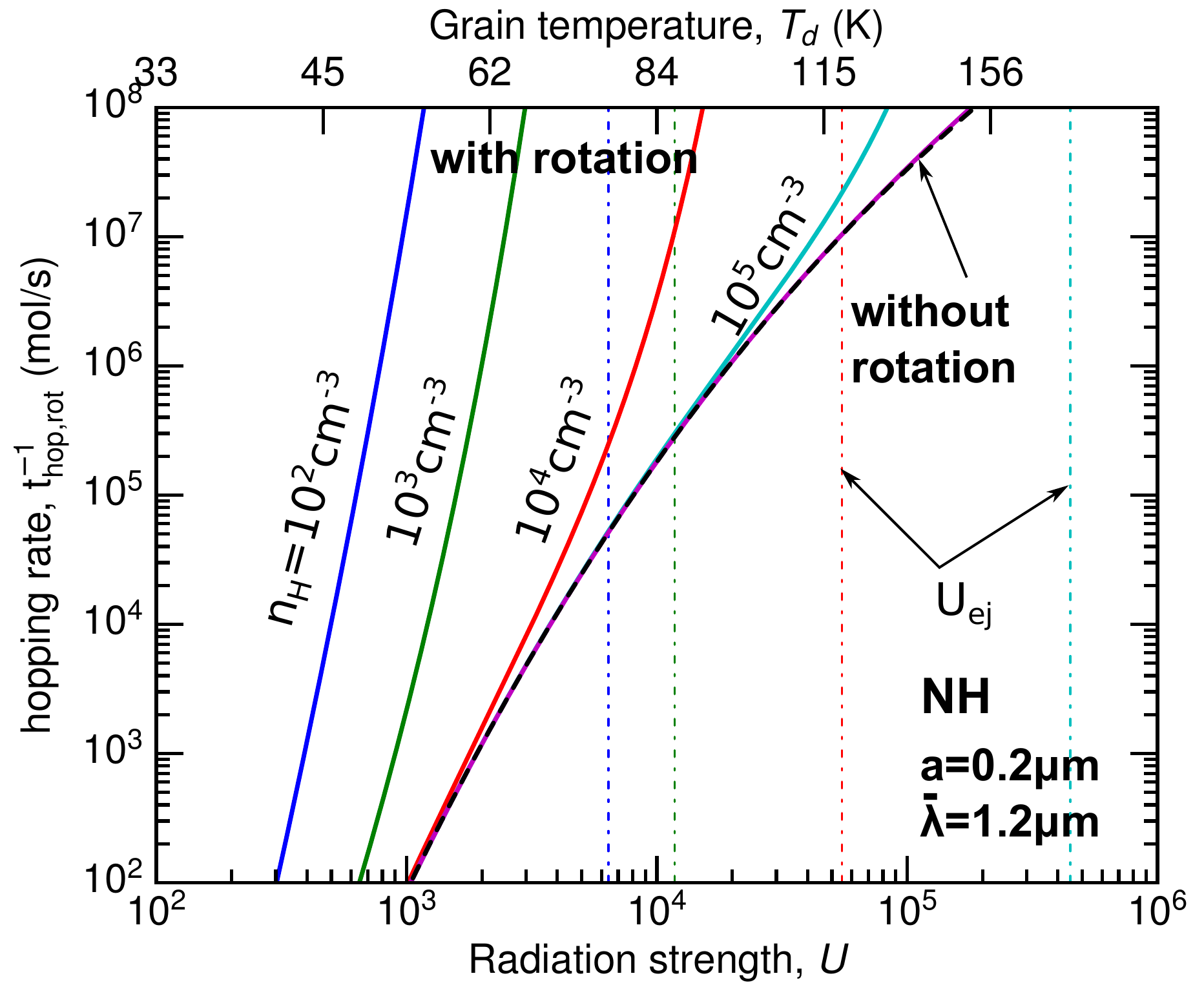}
\includegraphics[scale=0.5]{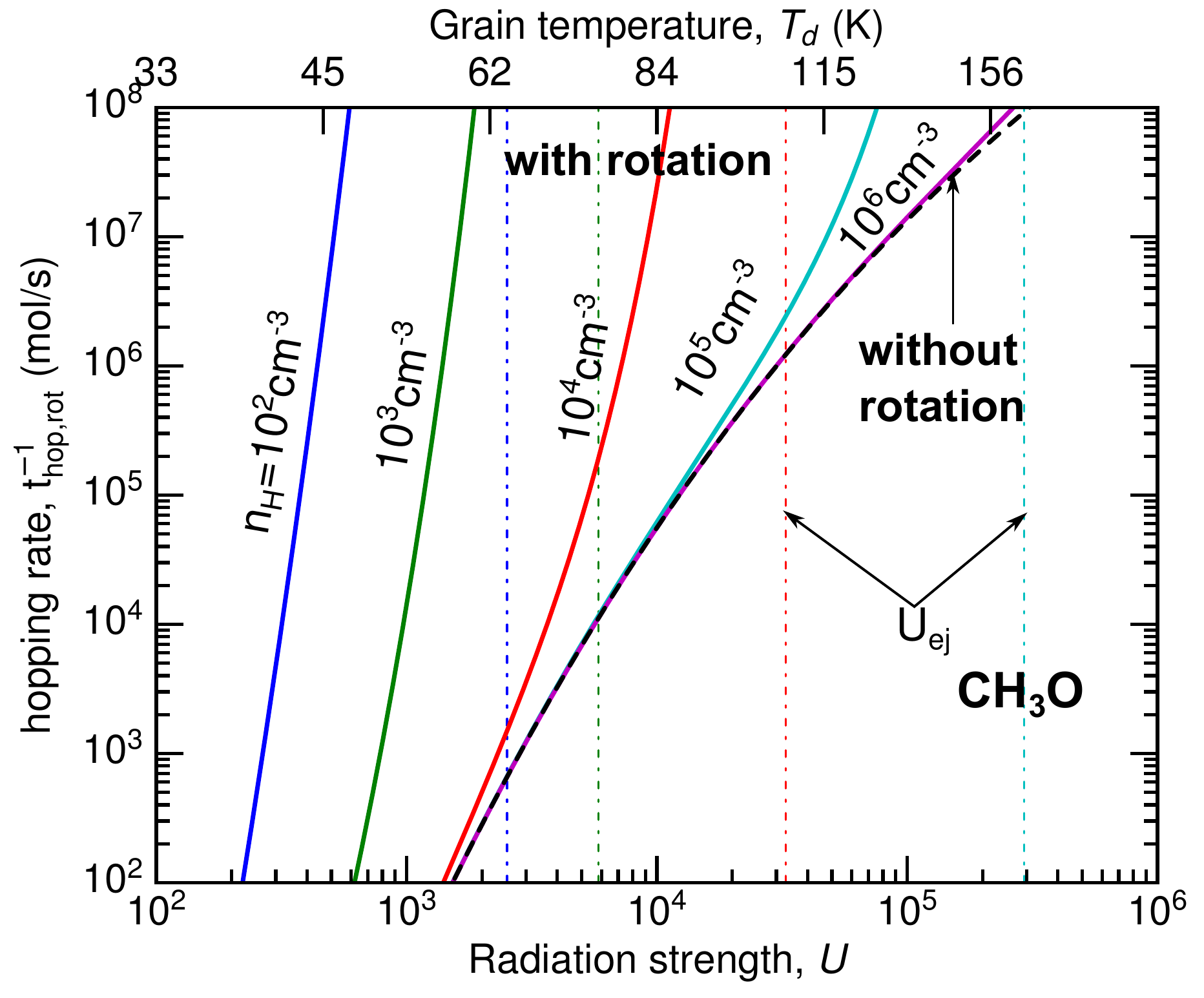}
\includegraphics[scale=0.5]{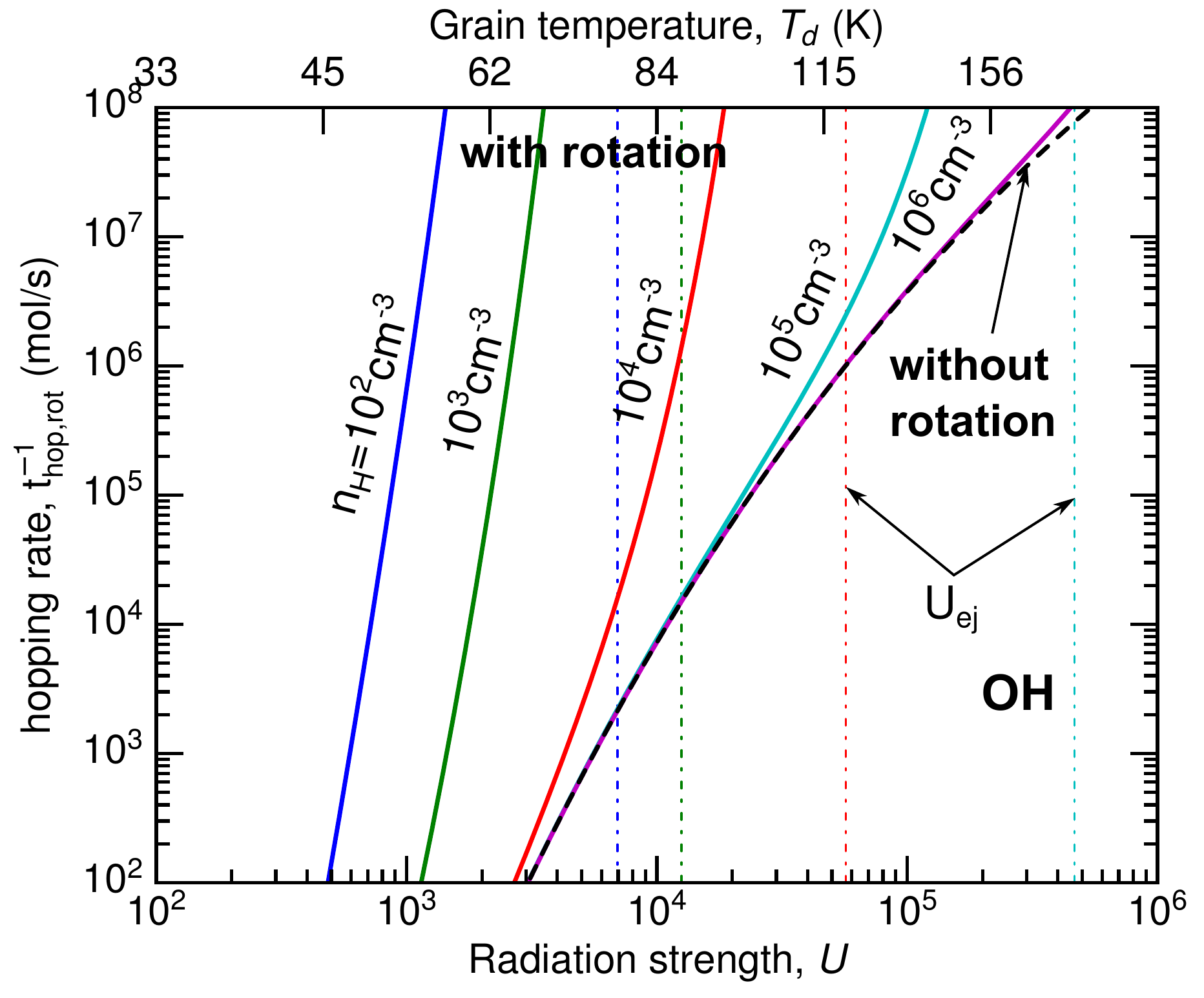}
\includegraphics[scale=0.5]{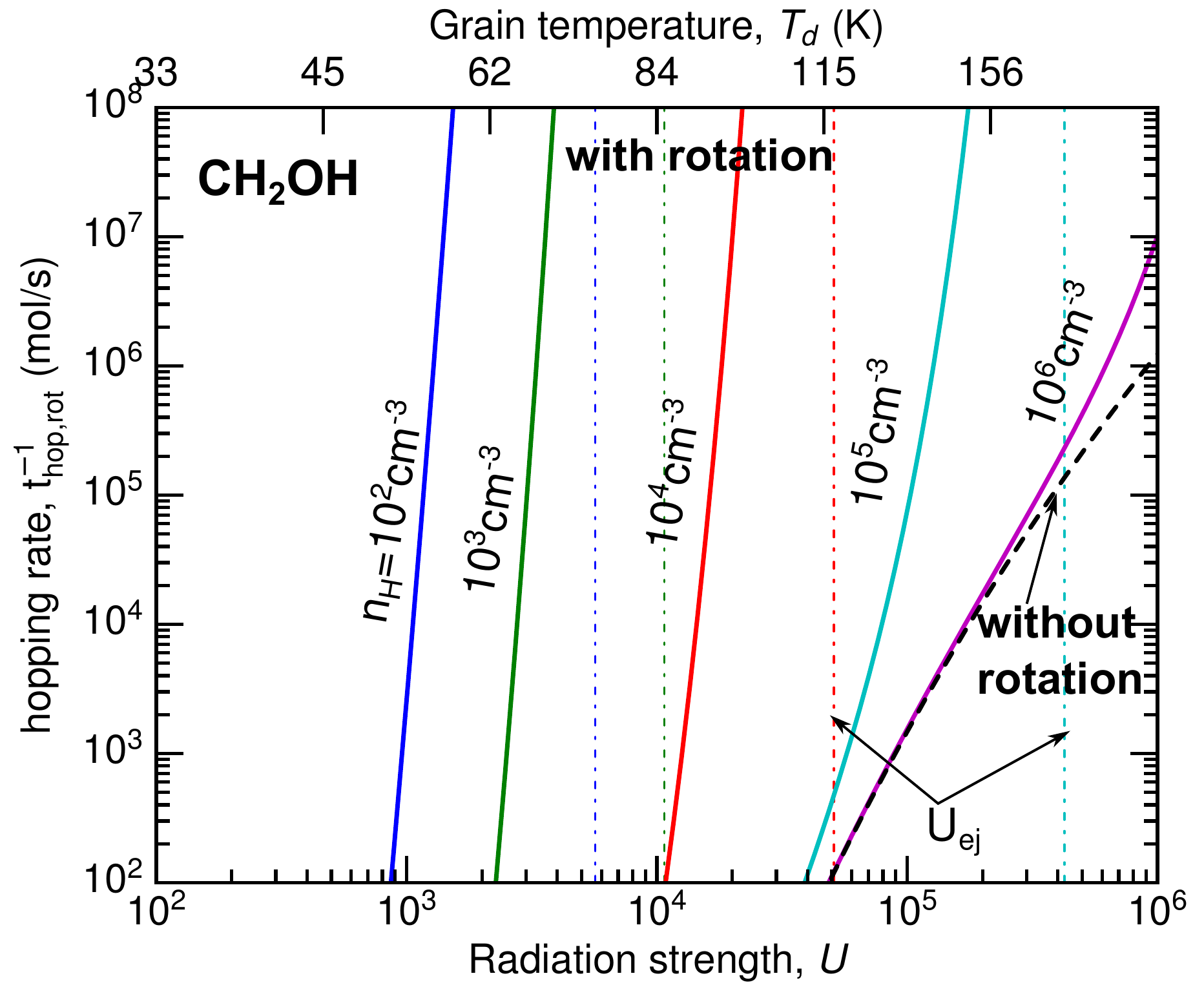}

\caption{Rate of ro-thermal hopping of primary radicals as a function of the radiation intensity for different molecules, assuming the gas density $n_{\H}=10^{2}-10^{6}\cm^{-3}$, $a=0.2\mum$ and $\bar{\lambda}=1.2\mum$. Ro-thermal hopping is faster than thermal hopping due to grain suprathermal rotation. The vertical lines show the threshold of the radiation strength at which the direct ejection of molecules by centrifugal force occurs.}
\label{fig:khop}
\end{figure*}

Figure \ref{fig:khop} shows the rate of ro-thermal hopping of primary radicals as a function of the incident radiation strength $U$ for the different molecules from simple radical to more complex molecules for the different gas densities $n_{\H}$. The grain temperature is shown in the upper horizontal axis for convenience. The rate of thermal hopping (without rotation) is also shown for comparison. We can see that the ro-thermal hopping depends closely on the local gas density, whereas thermal hopping depends only on the local radiation strength (grain temperature). Ro-thermal hopping is always dominates over thermal hopping, and the effect increases with decreasing the gas density. Even at a high density of $n_{\H}=10^{6}\cm^{-3}$, the ro-thermal hopping is still dominant over thermal hopping for $U>3\times 10^{4}$ ($T_{d}> 100\K$).

\begin{figure*}
\includegraphics[scale=0.5]{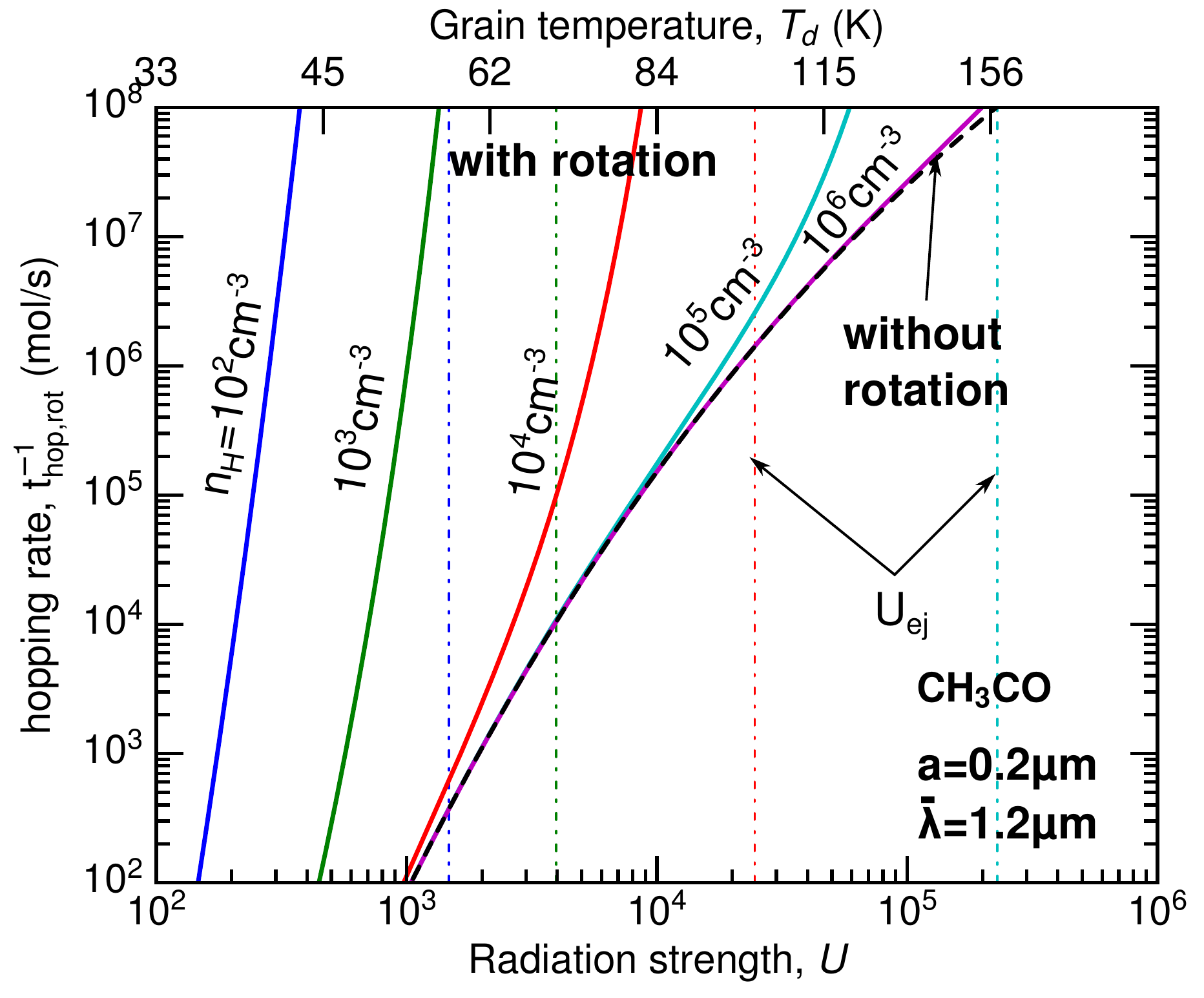}
\includegraphics[scale=0.5]{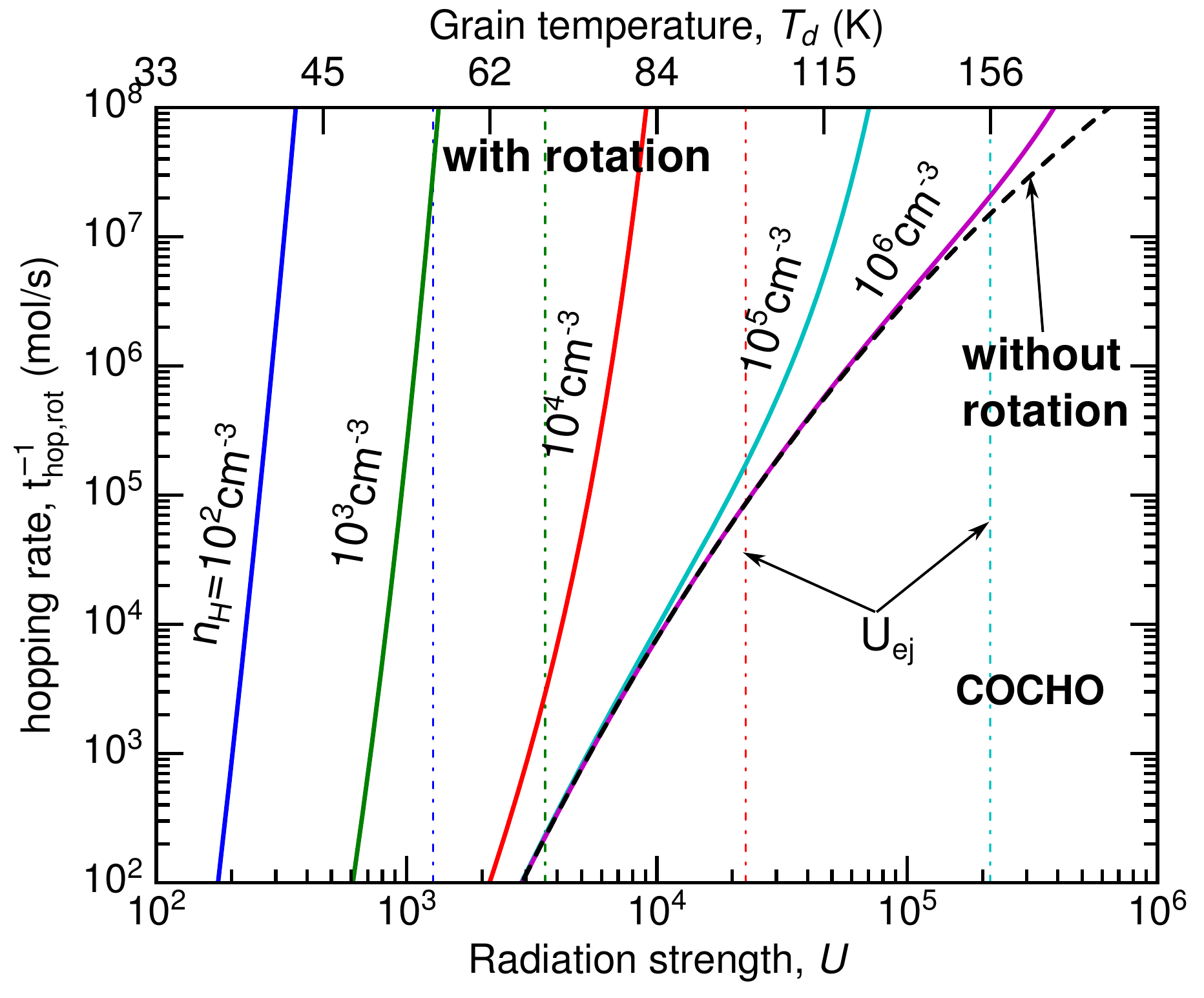}
\includegraphics[scale=0.5]{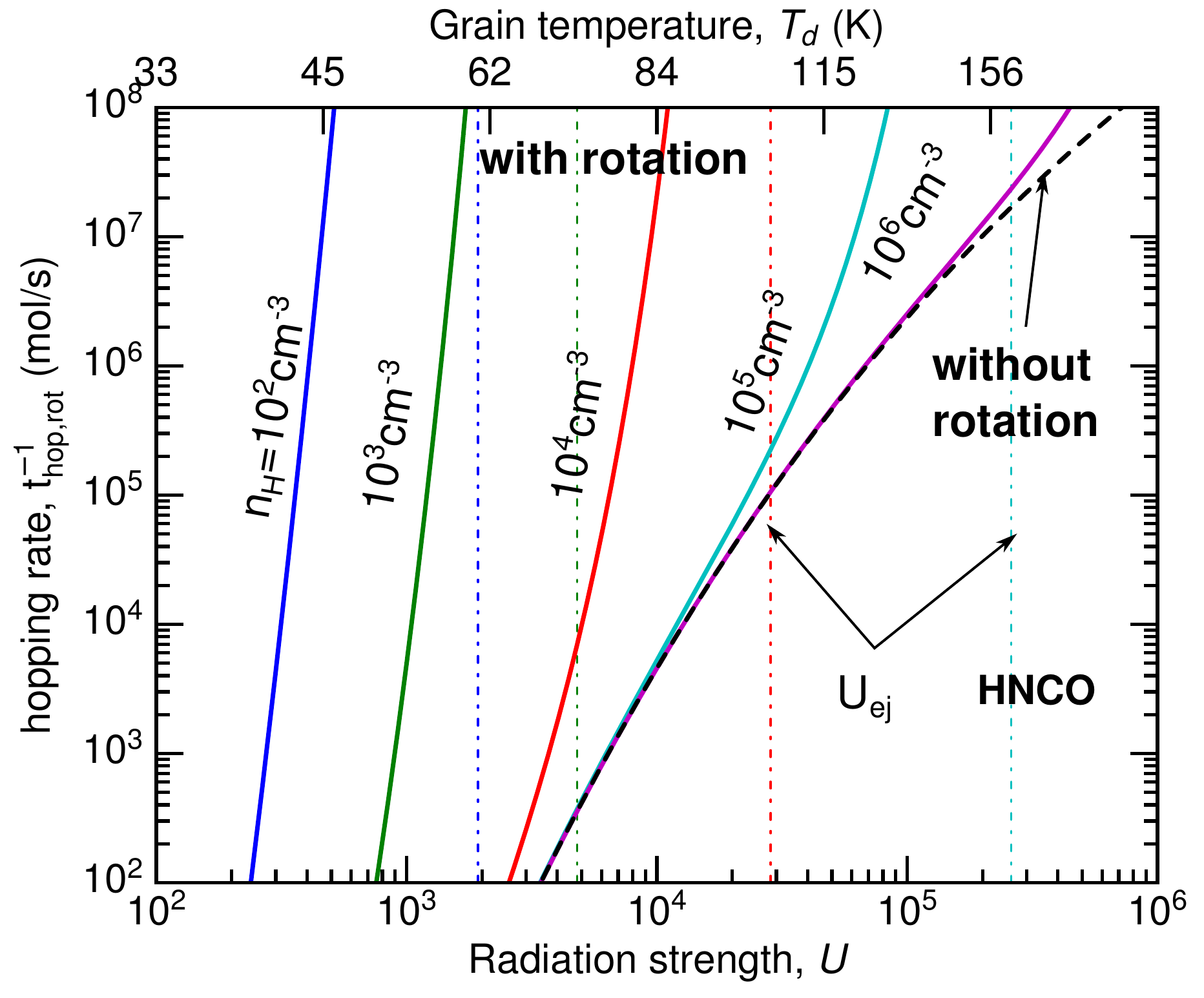}
\includegraphics[scale=0.5]{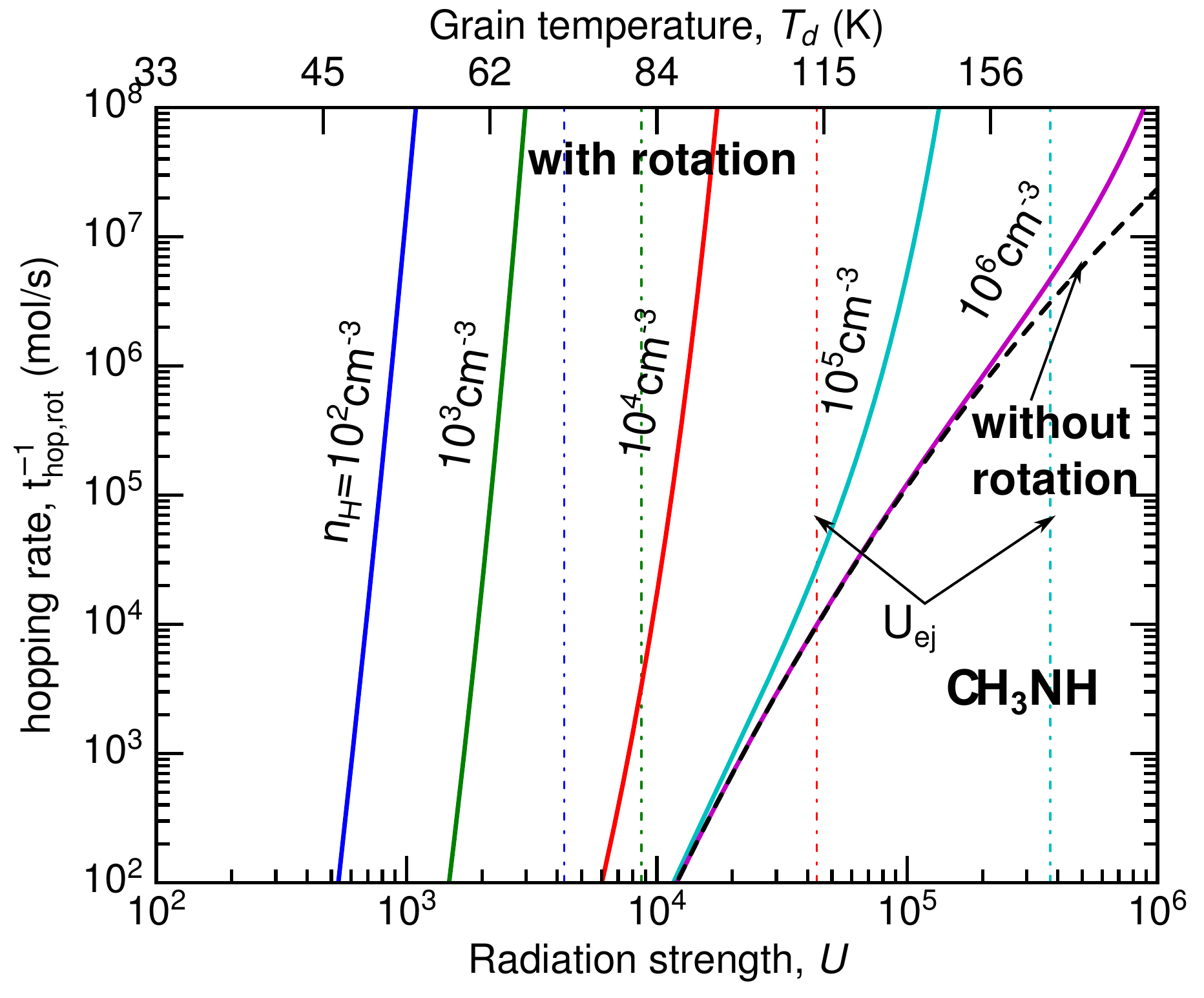}
\caption{Same as Figure \ref{fig:khop} but for secondary radicals. Ro-thermal hopping is faster than thermal hopping due to grain suprathermal rotation.}
\label{fig:khop2}
\end{figure*}

Figure \ref{fig:khop2} shows the rate of ro-thermal hopping as a function of the radiation strength for secondary radicals. The similar trend is observed where the ro-thermal hopping rate is much larger than that of thermal hopping.

We are interested in popular molecules detected in hot cores, including formic acid (HCOOH), methyl-formate (CH$_{3}$COOH), ethanol (C$_2$H$_5$OH), dymethyl ether (CH$_3$OCH$_3$), acetone (CH$_{3}$COCH$_{3}$), ethylene-glycol (CH$_{2}$OH)$_{2}$, and glycoaldehyde (CH$_2$OHCHO). These complex molecules may be formed via the following chemical reactions (see \citealt{2006A&A...457..927G}):
\bea
\OH +\H_{2}\O \rightarrow \H\CO\OH,\label{eq:HCOOH}\\
\OH + \CH_{3}\CO \rightarrow \CH_{3}\CO\OH,\\
\CH_{3}+ \CH_{2}\OH \rightarrow \C_{2}\H_{5}\OH,\\
\CH_{3} + \CH_{3}\O \rightarrow \CH_{3}\CO\CH_{3},\\
\OH + \CH_{2}\OH \rightarrow (\CH_{2}\OH)_{2},\\
\HCO + \CH_{3}\CO \rightarrow \CH_{2}\OH\CHO,\label{eq:CH2OHCHO}
\ena

\begin{figure*}
\includegraphics[scale=0.5]{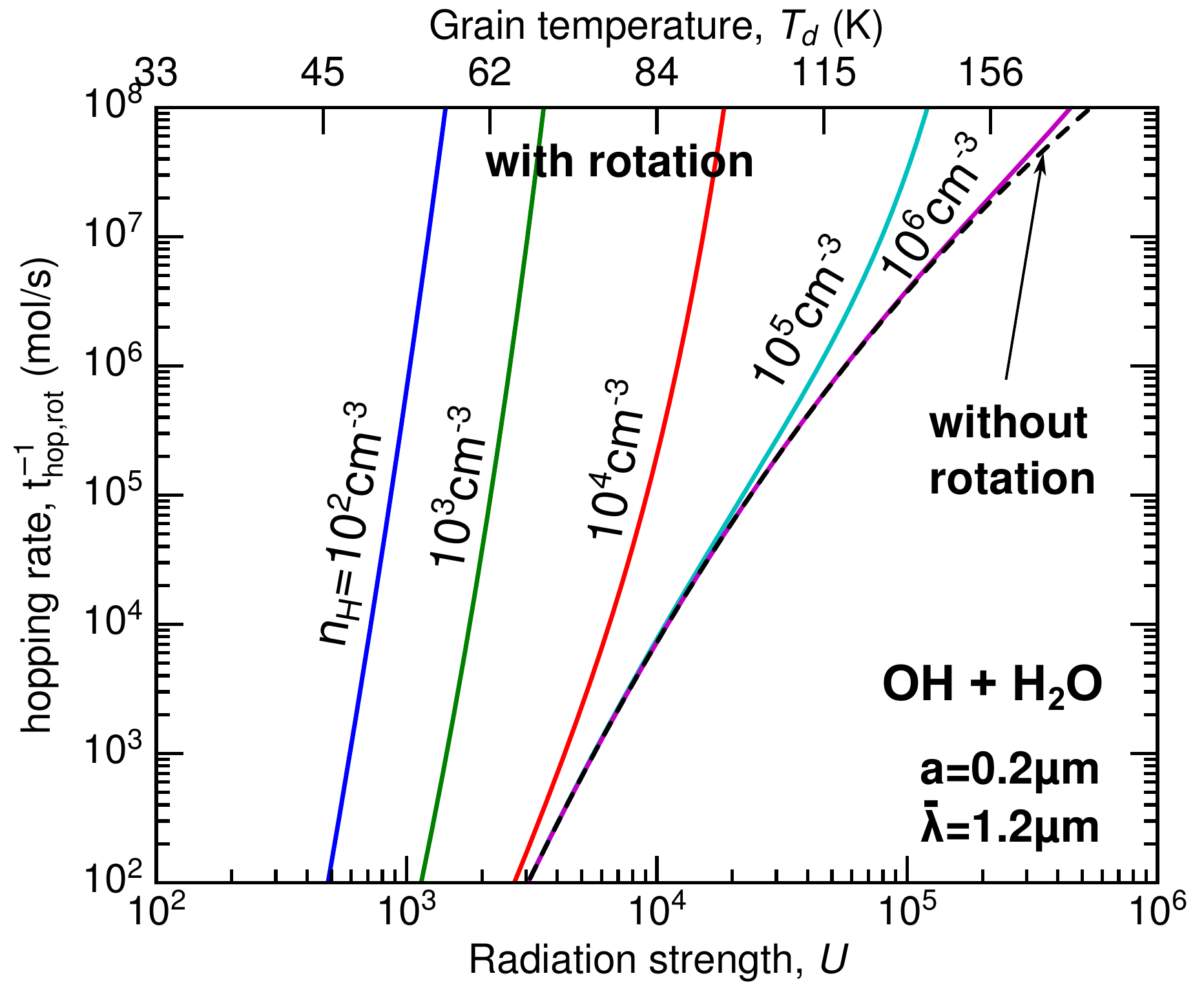}
\includegraphics[scale=0.5]{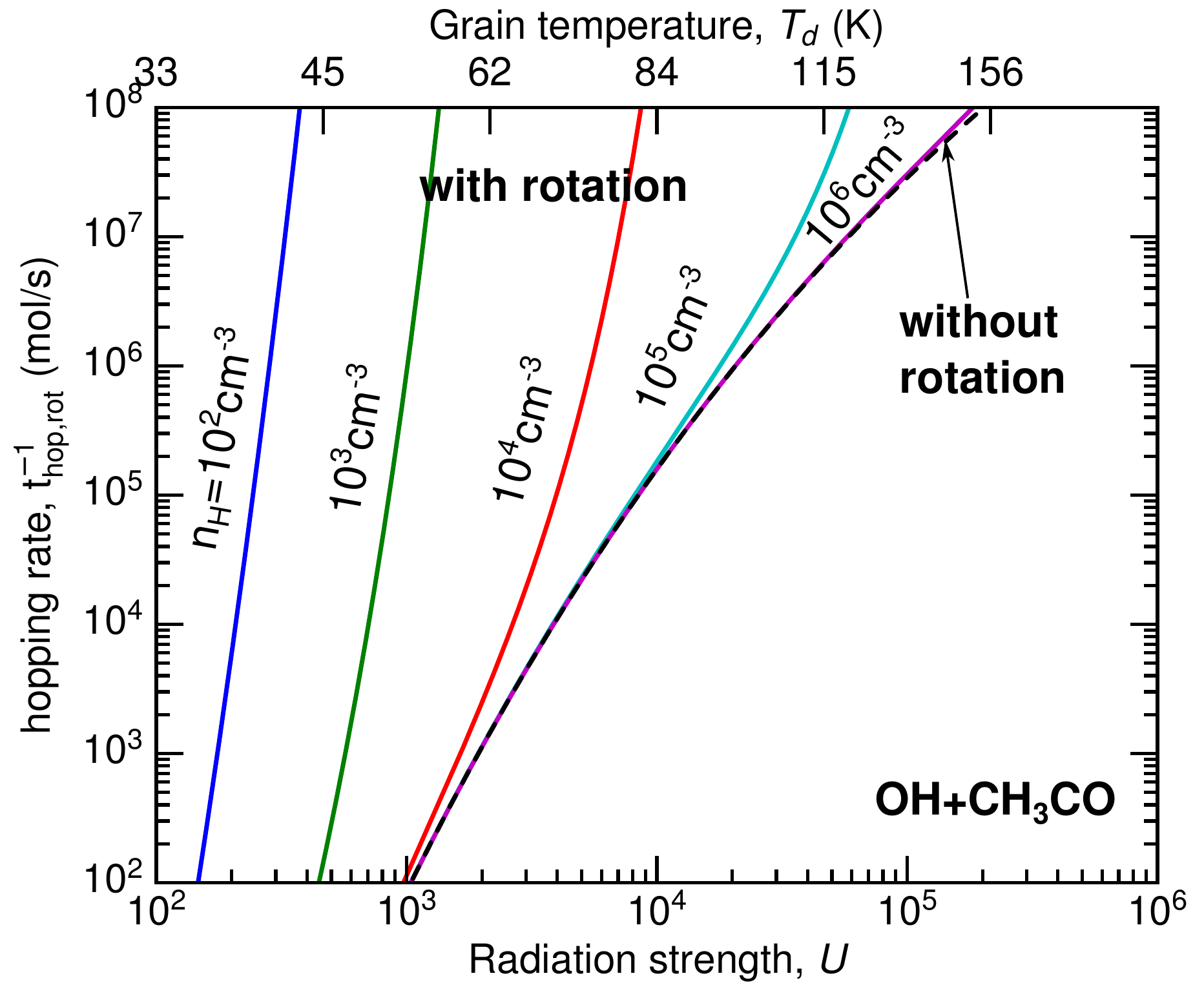}
\includegraphics[scale=0.5]{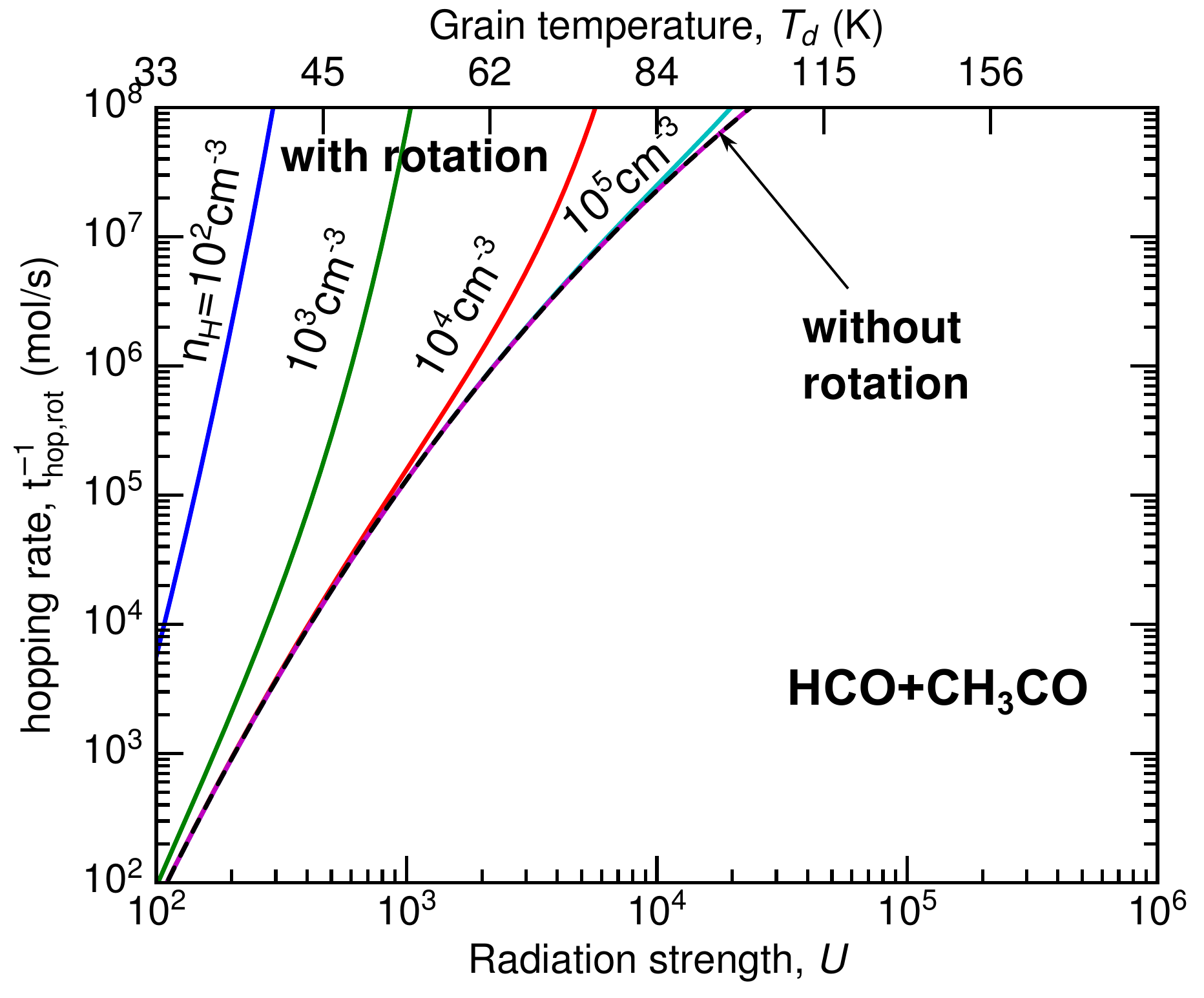}
\includegraphics[scale=0.5]{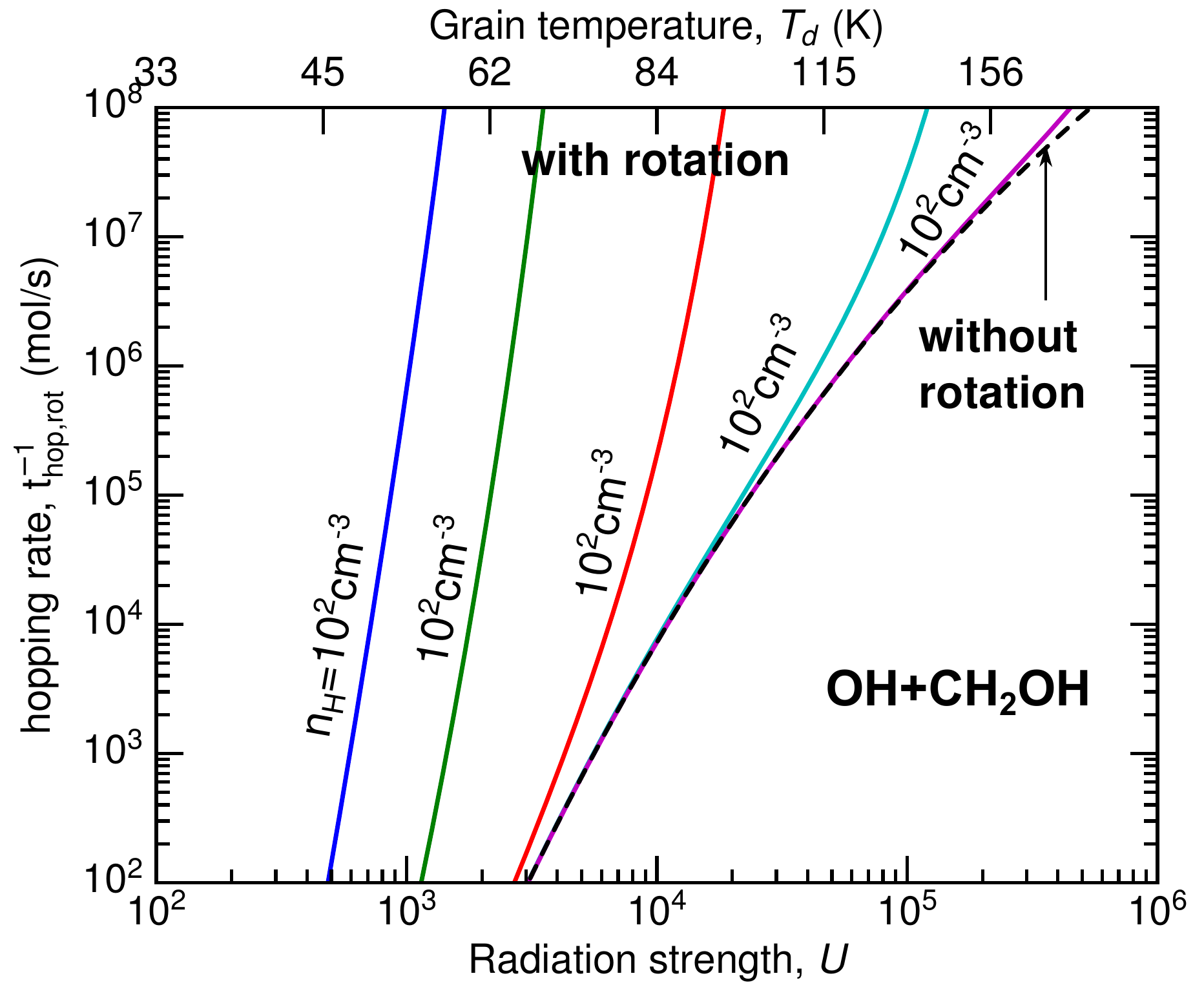}
\caption{Rate of chemical reaction due to ro-thermal hopping between primary radical (A) and secondary radical (B), $t_{hop,rot}^{-1}=t_{\rm hop,rot}^{-1}(A)+t_{\rm hop,rot}^{-1}(B)$, as a function of the radiation intensity for the different molecules, assuming the different local gas densities. Ro-thermal hopping is faster than thermal hopping due to grain suprathermal rotation.}
\label{fig:khop_AB}
\end{figure*}

Figure \ref{fig:khop_AB} shows the reaction rate of selective radicals given by Equations (\ref{eq:HCOOH})-(\ref{eq:CH2OHCHO}. As we see, at the same temperature of $T_{d}\sim 50$ K, the reaction rate of rotating grains is much larger than that of grains at rest if the local gas density is $n_{\H}<10^{5}\cm^{-3}$. Only in dense regions but low radiation field (i.e., $n_{\H}\ge 10^{5}, T_{d}<50\K$), the effect of ro-thermal hopping is negligible.

\subsubsection{Temperature thresholds of ro-thermal hopping}
We first calculate the temperature threshold of thermal hopping $T_{hop,0}=E_{\rm diff}/20$ for which by the hopping rate $t_{hop,0}^{-1}\sim 10^{4} \s^{-1}$ for $\nu\sim 5\times 10^{12} \s^{-1}$. These rates are chosen such that adsorbed species can scan the entire grain surface during a time of $10^{4}$yr, which is comparable to the evolution stage of star-forming regions. The temperature threshold of ro-thermal hopping which is required to achieve these rates is given by Equation (\ref{eq:dThop}).

\begin{figure*}
\includegraphics[width=0.5\textwidth]{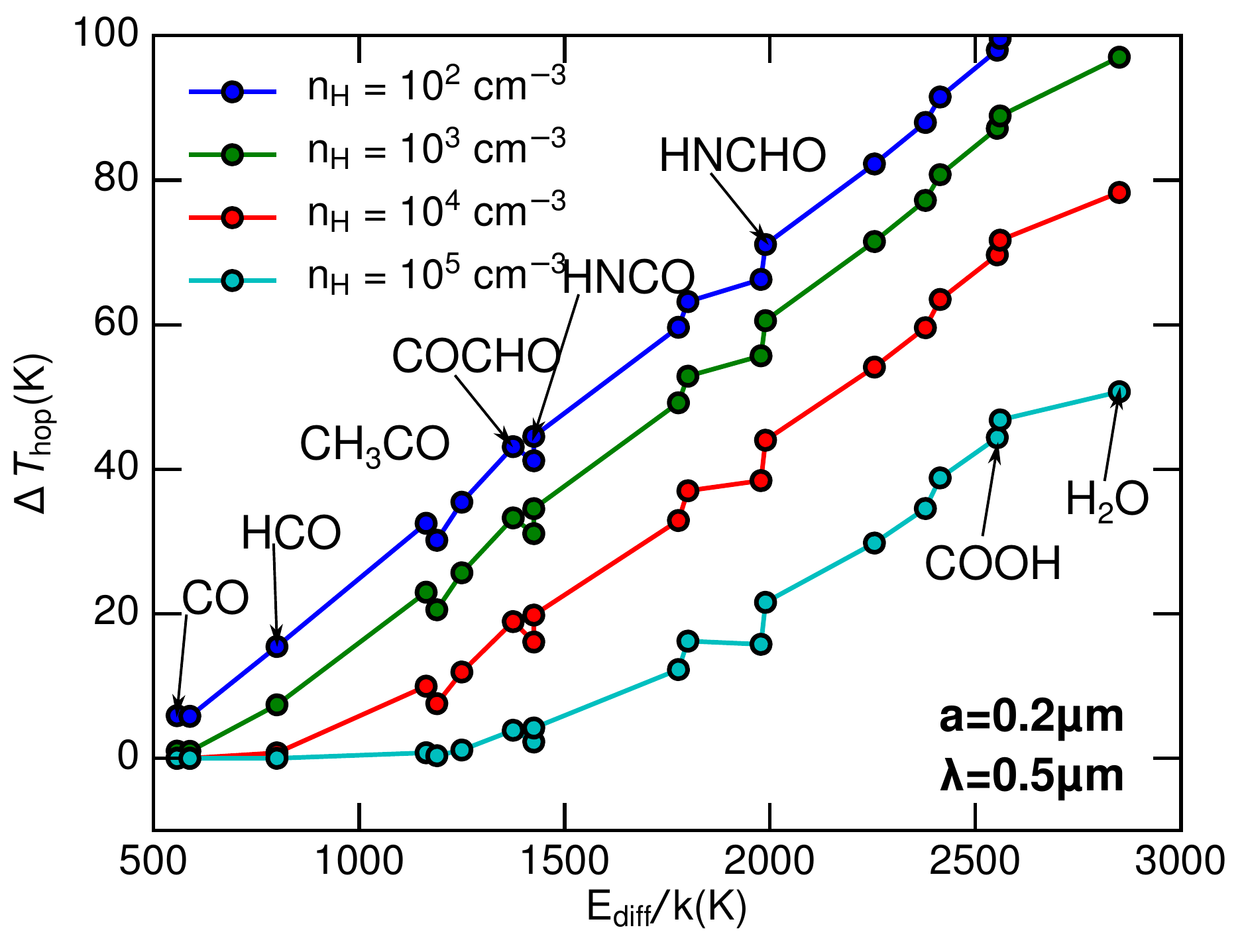}
\includegraphics[width=0.5\textwidth]{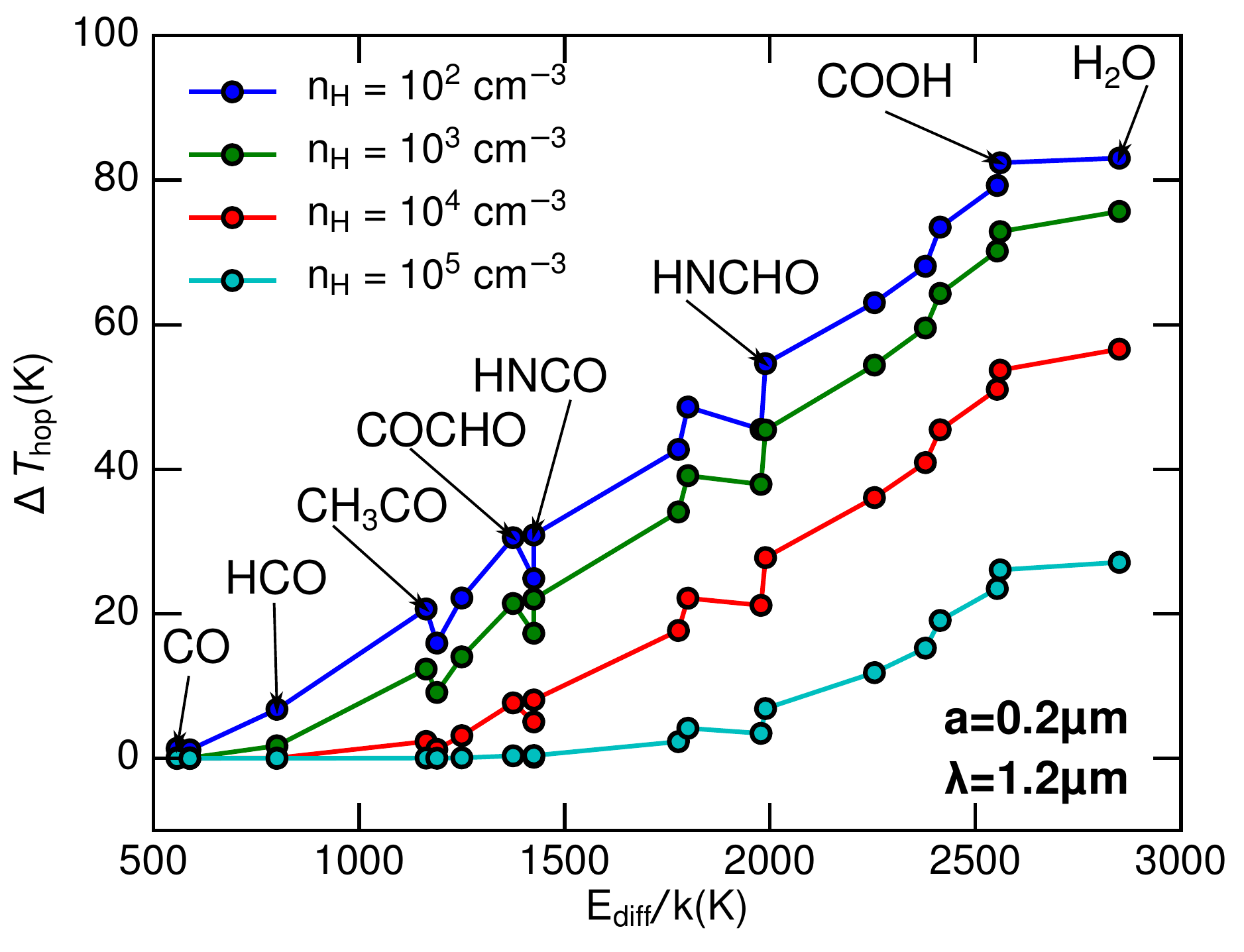}
\includegraphics[width=0.5\textwidth]{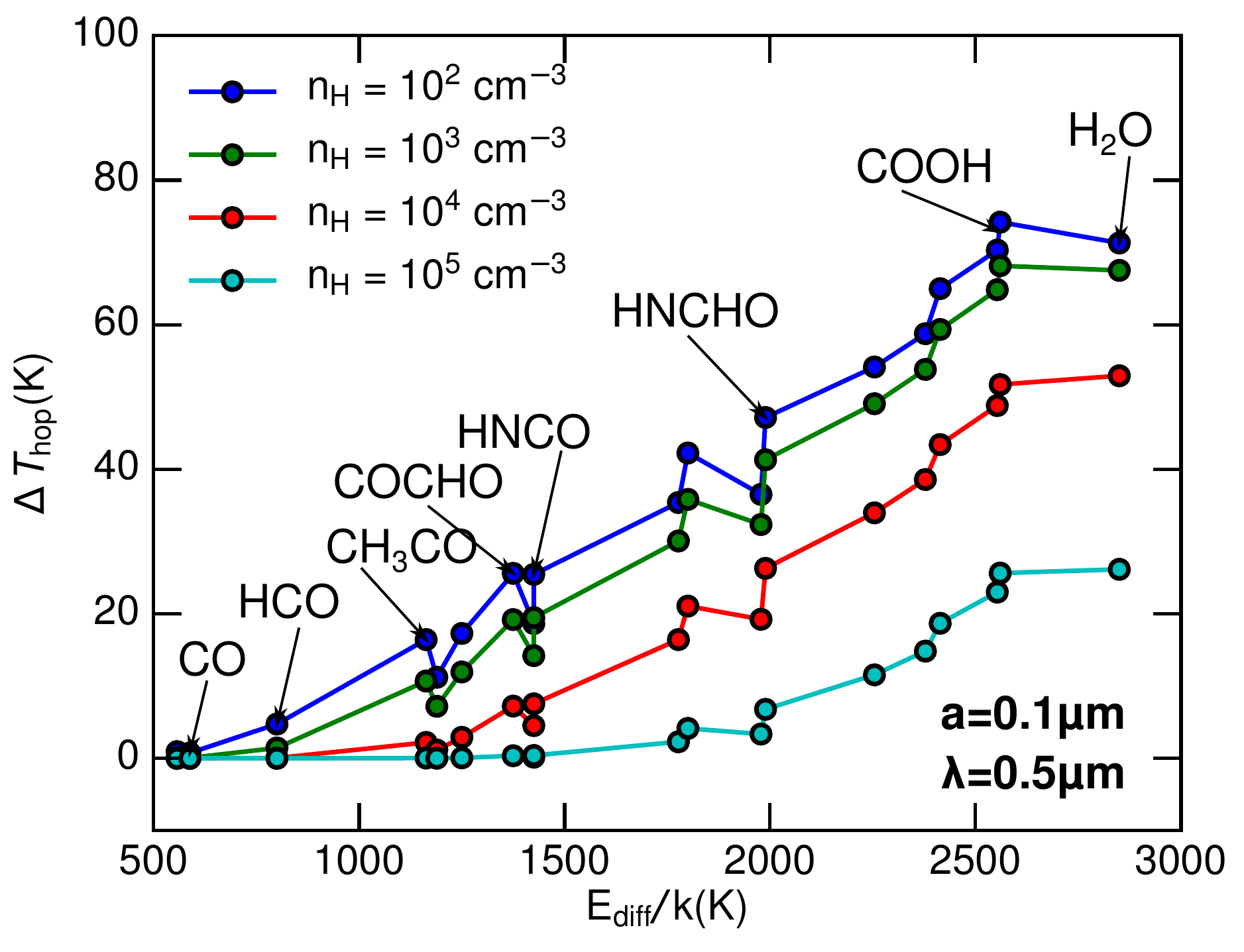}
\includegraphics[width=0.5\textwidth]{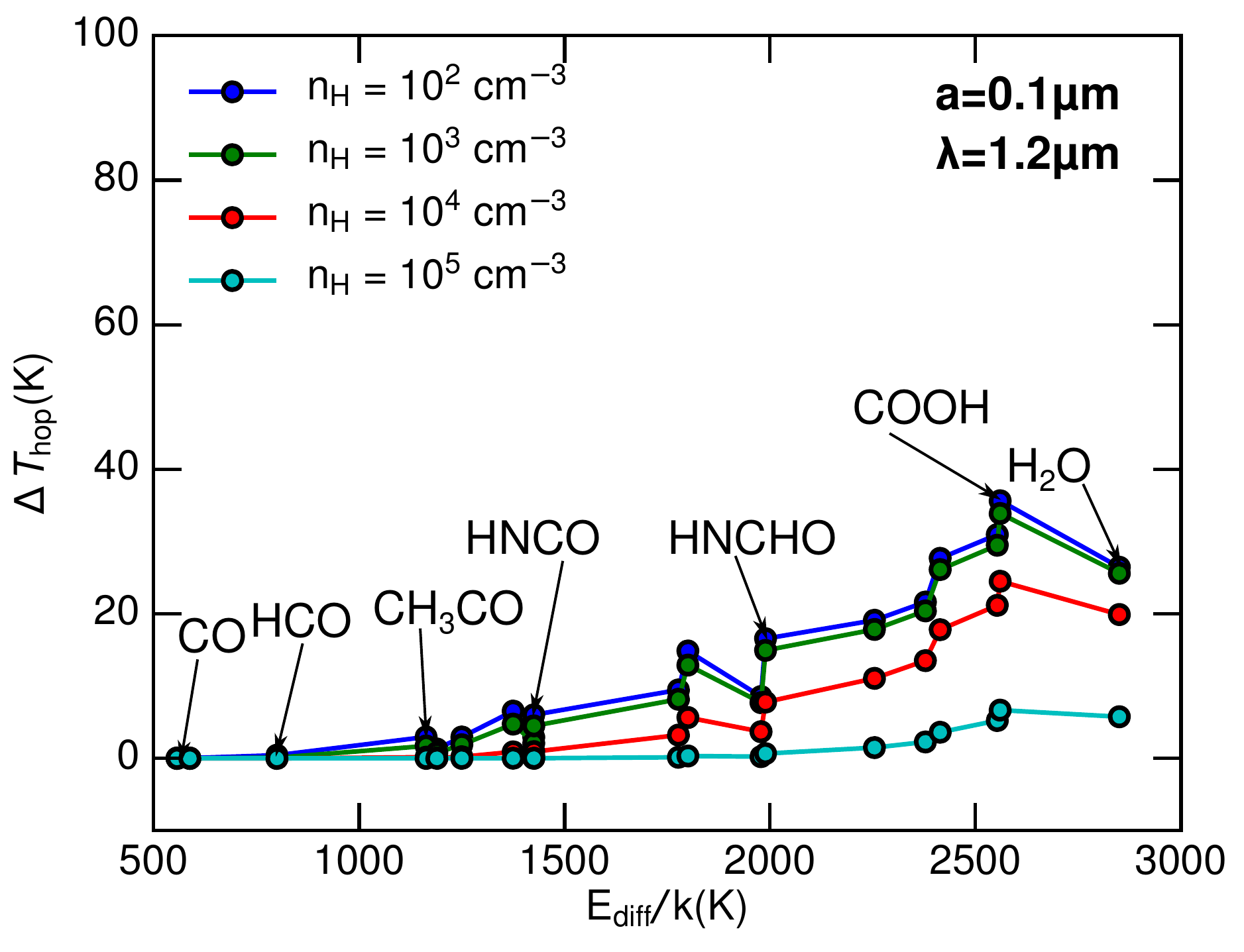}

\caption{Difference in the temperature threshold of ro-thermal hopping and that of thermal hopping, $\Delta T=|T_{\rm hop,rot}-T_{\rm hop,0}$, as a function of molecule binding energy for $a=0.1\mum$ and $a=0.2\mum$ assuming the radiation spectrum of $\bar{\lambda}=0.5, 1.2\mum$. The difference increases with increasing $E_{\rm diff}$ and with decreasing the gas density.}
\label{fig:deltaT}
\end{figure*}

Figure \ref{fig:deltaT} shows $\Delta T=|T_{\rm hop,rot}-T_{\rm hop,0}|$ as a function of the diffusion energy ($E_{\rm diff})$ for the different molecules, assuming the different gas densities. The effect of ro-thermal hopping is more important for lower gas densities. Ro-thermal hopping is also more efficient for adsorbed molecules with higher binding energy where ro-thermal hopping can occur at more than 100 K lower than the thermal hopping. The efficiency of ro-thermal hoping is more efficient for stellar photons of $\bar{\lambda}=0.5\mum$ but less efficient for reddened photons with $\bar{\lambda}=1.2\mum$ because the grain rotation rate from RATs decreases with the wavelength (see Eq. \ref{eq:omega_RAT1}). Even at high gas densities of $n_{\rm H}=10^{5}\cm^{-3}$, ro-thermal hopping can still occur at temperatures lower than thermal hopping.

\section{Discussion}\label{sec:discuss}
\subsection{Importance of grain suprathermal rotation for surface astrochemistry}
Grain surfaces play a key role for the formation and desorption of complex molecules (e.g., COMs) in astrophysical environments. To form molecules, adsorbed species must migrate between binding sites on the grain surface. Subsequently, resulting new molecules must be desorbed from the ice mantle to the gas phase. Thermal hopping is widely thought to be an important mechanism for the mobility of molecules on the ice mantle in star-forming regions where grains are warmed up by stellar radiation.

In this paper, we showed that, subject to strong stellar radiation, grains can be spun-up to suprathermal rotation, such that the energy diffusion barrier of adsorbed species is reduced due to the contribution of centrifugal energy. As a result, the rate of thermal hopping on suprathermally rotating grain surfaces is much higher than that on grain surfaces at rest. We term this mechanism {\it ro-thermal hopping}. Combining the increased ro-thermal hopping with ro-thermal desorption (\citealt{Hoang:2019wra}) due to grain suprathermal rotation, the rate of formation of molecules on rotating surfaces is found to be much higher than predicted by thermal hopping on non-rotating grains.

We note that, in the current paradigm of surface astrochemistry, the temperature of grain surfaces heated by local radiation fields controls the mobility of absorbed species, formation, and desorption of molecules. Here we found that the gas density $n_{\H}$ is a crucially important physical parameter because it determines the ultimate energy barrier that adsorbed species must overcome as due to the effect of centrifugal energy induced by RATs.

\subsection{Implication for segregation of ice mixtures}
Infrared observations toward cold pre-stellar cores show that most CO$_2$ ice is mixed with $\H_{2}\O$ ice, and CO$_2$ and CO ice are the two separate layers. Therefore, CO$_2$ and $\H_{2}\O$ ice are thought to form simultanouesly and exist in the mixture of CO$_2$-H$_2$O ices (see \citealt{Fayolle:2011eo}). Yet, observations show the segregation of CO$_{2}$  from $\H_{2}\O$ ices toward protostellar regions (\citealt{2008ApJ...678.1005P}). The authors suggested that the segregation of CO$_2$-$\H_{2}\O$ ices arises from radiative heating of the ice mantle to above 50 K by protostars that causes the diffusion of CO$_2$ molecules in the ice mantle. Experiments in \cite{Hodyss:2008cg} observed the onset of the segregation in H$_2$O:CO$_2$ ice mixtures at temperatures of $T_{d}\sim 60\K$. Experiments by \cite{Oberg:2009dm} observed ice segregation at $T_{d}\sim 40-60\K$, which is explained by means of thermal diffusion. 

Thermal processing is thought to play a key role in segregation and crystallization of ices (see \citealt{Boogert:2015fx} for a review). The rate ro-thermal diffusion (i.e., segregation) of molecules in the ice mantle is described by the same diffusion equation where $E_{\rm diff}$ is replaced with the segregation energy $E_{\rm seg}$:
\bea
k_{\rm seg,rot}=k_{\rm seg,0}\exp\left(\frac{m\langle \phi_{\rm cen}\rangle}{kT_{d}} \right),
\ena
where
\bea
k_{\rm seg,0}=\nu_{0}\exp\left(\frac{E_{\rm seg}}{kT_{d}} \right).
\ena

The energy barrier for the segregation is uncertain, but it is expected to be higher than the diffusion barrier. Here one adopts $E_{\rm seg}= 0.7E_{b}$ as in \cite{2013ApJ...765...60G}.

\begin{figure*}
\includegraphics[scale=0.45]{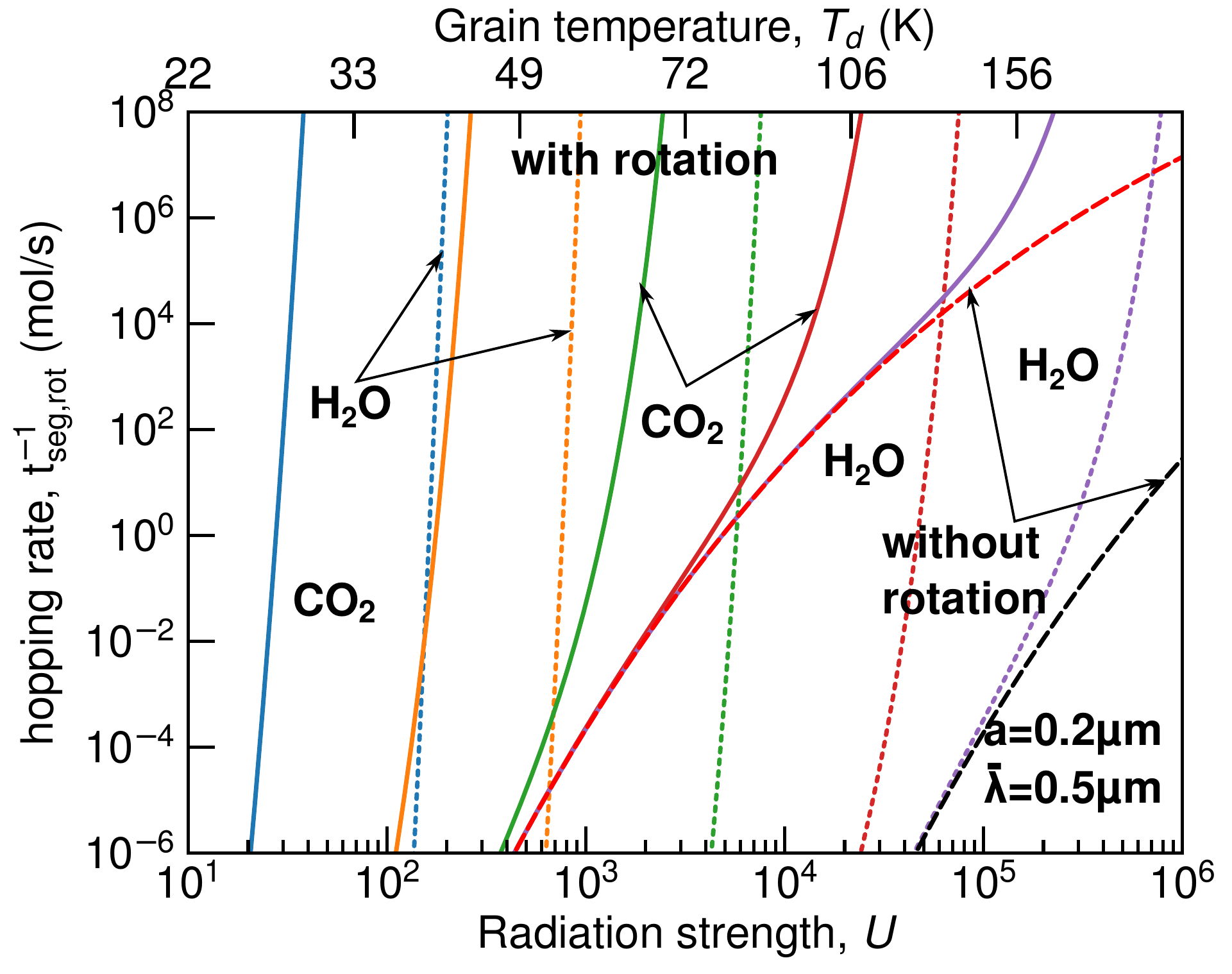}
\includegraphics[scale=0.45]{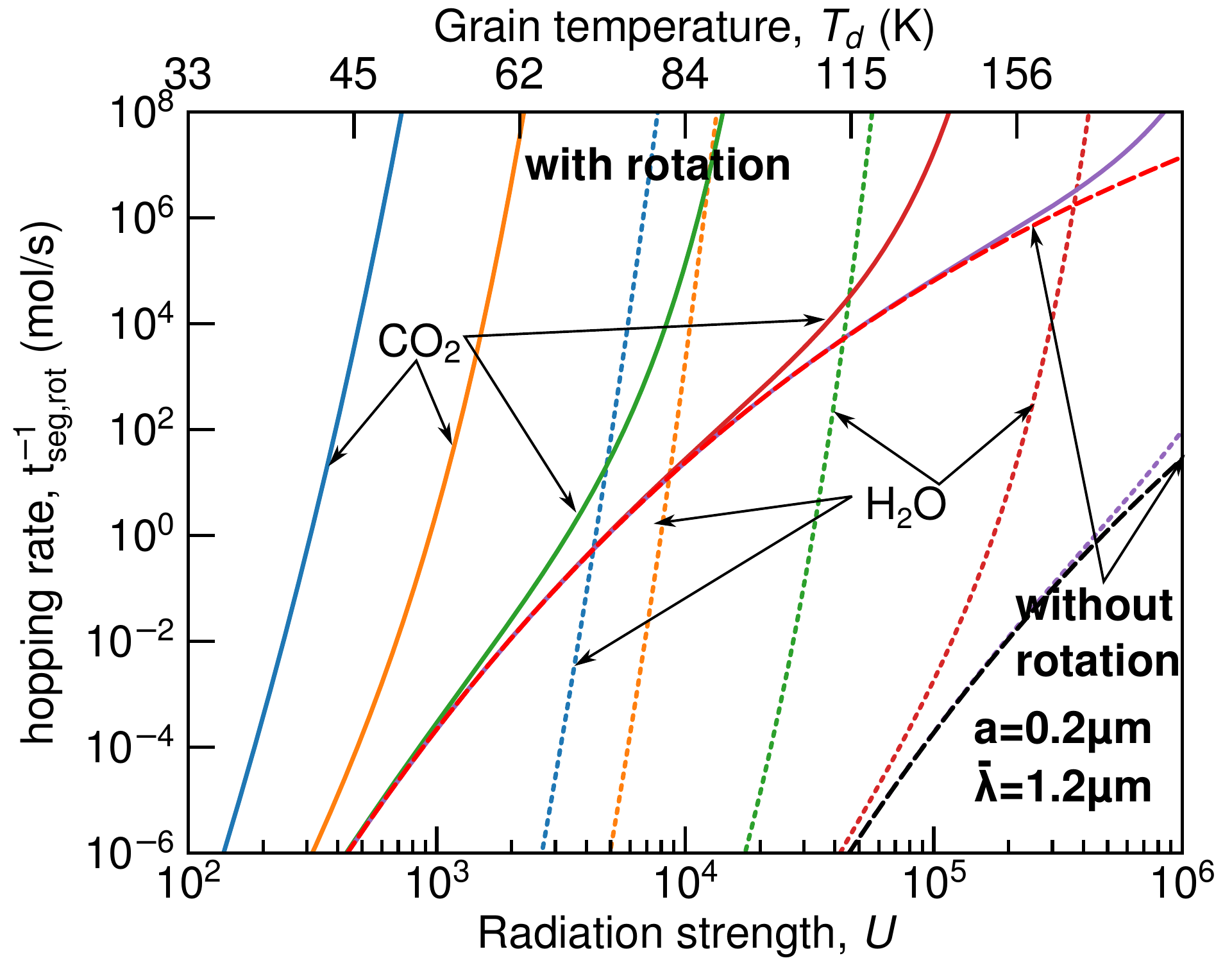}
\caption{Rate of ro-thermal segregation vs. thermal segregation of CO$_2$ (solid lines) and H$_{2}$O (dotted lines) for the different gas densities.}
\label{fig:seg}
\end{figure*}

In Figure \ref{fig:seg}, we show the rate of ro-thermal segregation for $\H_{2}\O$ and $\C\O_{2}$ in the icy grain mantle of size $a=0.2\mum$, assuming the radiation spectrum of $\bar{\lambda}=0.5\mum$ (left panel) and $\bar{\lambda}=1.2\mum$ (right panel). First, for a given grain temperature, the diffusion rate of CO$_2$ is much larger than that of $\H_{2}\O$, which implies the segregation of these two species over time. Second, thermal diffusion of CO$_2$ is negligible ($k_{seg,0}<1$) when the grain temperature is $T_{d}<80\K$, but the rate of ro-thermal segregation can be efficient even at low temperatures of $T_{d}<50\K$ for $n_{\H}<10^{4}\cm^{-3}$. Moreover, since $m(\CO_2)\approx 2.45m(\H_{2}\O)$, the rate of ro-thermal segregation of CO$_2$ is larger than that of H$_2$O due to the increase of centrifugal energy with the molecule mass.

\begin{figure*}
\includegraphics[width=0.5\textwidth]{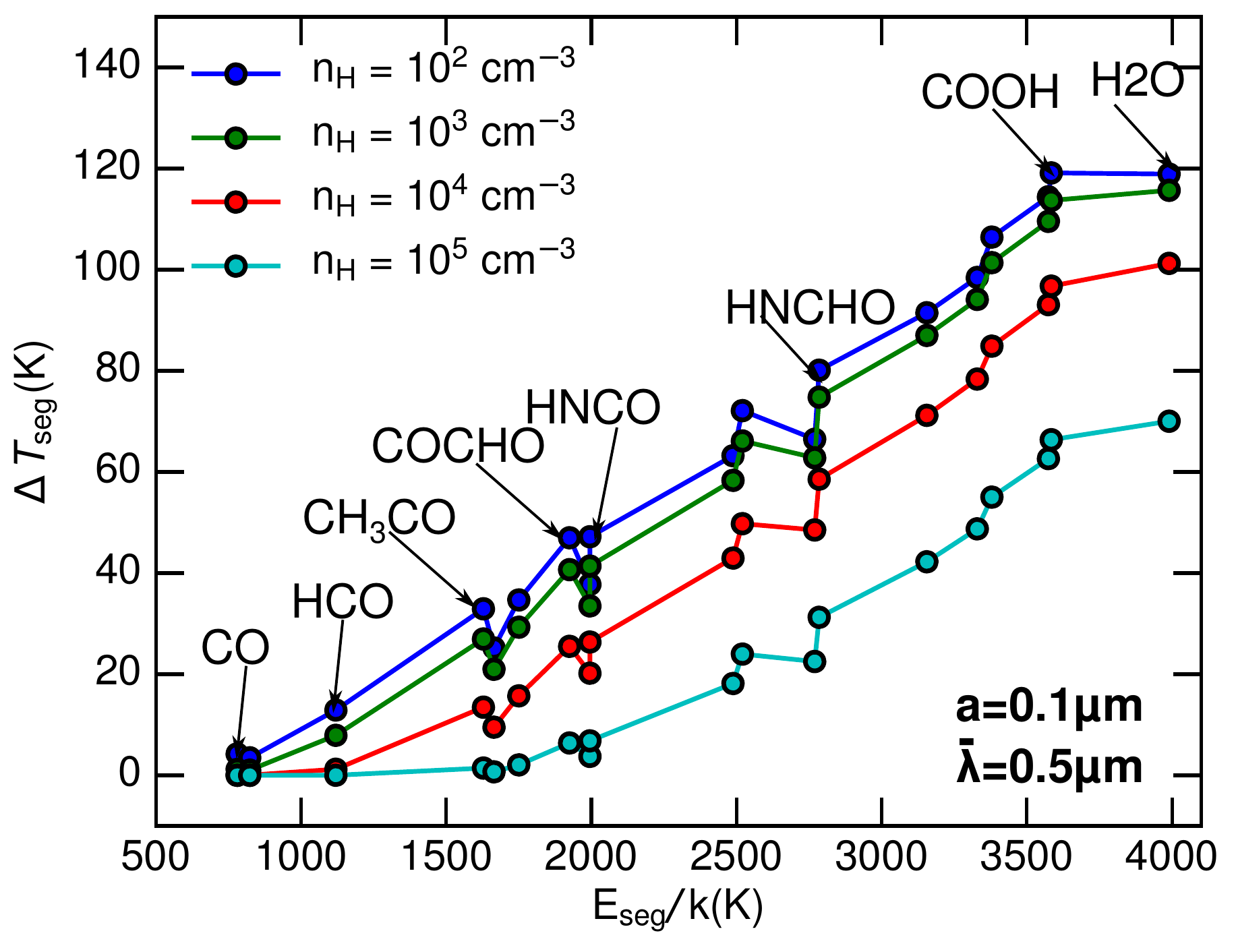}
\includegraphics[width=0.5\textwidth]{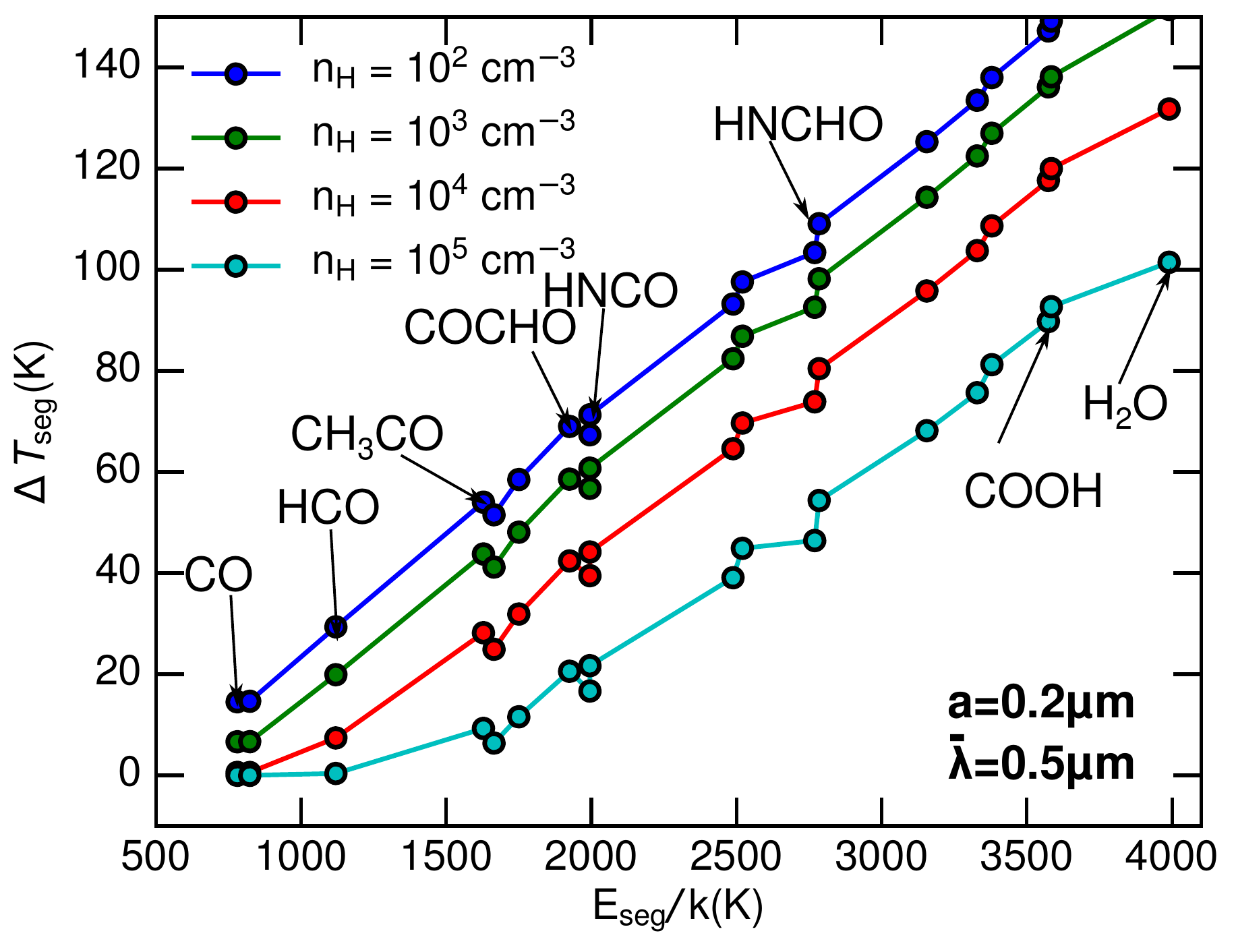}
\caption{Difference in the temperature threshold of ro-thermal segregation and that of thermal segregation, $\Delta T=|T_{\rm seg,rot}-T_{\rm seg,0}$, as a function of molecule binding energy for $a=0.1\mum$ and $a=0.2\mum$ for $\bar{\lambda}=0.5\mum$. The difference increases with increasing $E_{\rm seg}=0.7E_{\rm D}$ and with decreasing the gas density.}
\label{fig:deltaT_seg}
\end{figure*}

The ro-thermal segregation can induce the crystallization of CO$_2$ and H$_2$O ice at lower temperatures than obtained by experiments (\citealt{Hodyss:2008cg}; \citealt{Fayolle:2011eo}).

Future high resolution observations by James Webb Space Telescope (JWST) would be useful to quantify the location of ice segregation and understand the underlying physics (\citealt{Boogert:2015fx}).


\subsection{Expected environments for ro-thermal hopping}
From Figure \ref{fig:khop} it shows that the ro-thermal hopping is most important in regions where $U/n_{\H}>0.1$. Therefore, in PDRs where $G_{0}/n_{\H}\sim 0.1-1$ (see e.g., \citealt{Tielens:2007wo}; \citealt{1997ARA&A..35..179H}; \citealt{Esplugues:2016dk}), ro-thermal hopping appears to be important. Similarly, the ro-thermal hopping effect is expected to be important in regions with strong radiation fields such as hot cores around massive protostars (see \citealt{Hoang:2019td}), surface and intermediate layers of protoplanetary disks. Moreover, the structure of astrophysical environments is unlikely homogeneous and tends to have clumpy structures with the different densities. As a result, ro-thermal diffusion and segregation can then occur at lower temperature if the local density is low.

In the classical grain surface chemistry, the formation of COMs is controlled by the time of warming-up phase. If the time is short, it is not sufficient for molecules to form COMs via slow thermal hopping process. Our ro-thermal hopping and ro-thermal desorption can form COMs faster, which requires shorter warming-up phase. Therefore, the accurate physics of molecule formation is unique to constrain the rate of cloud collapse and star formation.

The effect of ro-thermal desorption (\citealt{Hoang:2019wra}) and rotational desorption (\citealt{Hoang:2019td}) can release saturated species such as H$_{2}$O, CH$_{3}$OH, NH$_{3}$, at temperatures below their sublimation limits. Those evaporating species can trigger rich gas-phase chemistry in warming-up phase (see e.g., \citealt{2013RvMP...85.1021T}).

Finally, we note that the ro-thermal hopping and segregation depend crucially on the grain angular velocity because $\phi_{\rm cen}\propto \omega^{2}/T_{d}$. Therefore, the effect can be efficient whenever grains can be spun-up to suprathermal rotation, such as by mechanical torques due to gas flows and irregular grains (\citealt{2007ApJ...669L..77L}; \citealt{2018ApJ...852..129H}) and pinwheel torques (\citealt{1979ApJ...231..404P}). In this regime, the grain suprathermal rotation dominates the mobility of adsorbed species, formation, segregation, and desorption of newly formed molecules even when the ice mantle is cold. A detailed study of this regime will be presented elsewhere.

\section{Summary}\label{sec:sum}
In the present paper, we have studied the effect of grain suprathermal rotation by radiative torques on hopping and segregation of molecules in the ice mantles. Our principal results are summarized as below:

\begin{enumerate}

\item We studied the effect of grain suprathermal rotation on the thermal hopping of adsorbed species on the grain surface. The thermal hopping in the presence of rotation is called ro-thermal hopping.

\item Using the rotation rate of grains driven by RATs due to anisotropic radiation, we found that the rate of ro-thermal hopping is much larger than that of thermal hopping. Thus, the effect of grain suprathermal rotation plays a vital role in the formation and desorption of more complex molecules in star-forming regions. 

\item We found that the mobility, formation, and desorption of molecules from ice mantles do not only depend on the grain temperature, but also on the local gas density because the grain suprathermal rotation is determined by the radiation intensity and rotational damping by gas collisions. Therefore, to achieve an accurate understanding on the physical and chemical properties using molecular tracers, one needs to look for detailed modeling of grain rotational dynamics and surface chemistry combined with observational data.

\item We studied the effect of grain suprathermal rotation on ice segregation and found that the segregation of CO$_2$ from H$_2$O ice mixtures can occur at lower temperatures than predicted by classical thermal diffusion, depending on the local gas density. We expect that grain suprathermal rotation also plays an important role in the crystallization of ice.

\end{enumerate}

\acknowledgments
I thank Professor Takashi Onaka for drawing my attention to the problem of ice segregation. This work was supported by the National Research Foundation of Korea (NRF) grants funded by the Korea government (MSIT) through the Basic Science Research Program (2017R1D1A1B03035359) and Mid-career Research Program (2019R1A2C1087045).


\bibliography{ms.bbl}

\end{document}